\documentclass[aps,
prb,
twocolumn,
nofootinbib,
]{revtex4-1}

\bibpunct{[}{]}{;}{n}{}{}

\usepackage{graphicx}
\usepackage{subfigure}
\usepackage[utf8]{inputenc}
\usepackage{amsmath}
\usepackage{amssymb}
\usepackage{appendix}
\usepackage{xcolor}

\usepackage{hyperref}
\usepackage{comment}
\usepackage{pdfpages}
\usepackage{cleveref}

\usepackage[normalem]{ulem}

\crefname{equation}{Eq.}{Eqs.}
\Crefname{equation}{Equation}{Equations}
\crefname{figure}{Fig.}{Figs.}
\Crefname{figure}{Figure}{Figures}
\crefname{section}{Sec.}{Secs.}
\Crefname{section}{Section}{Sections}
\crefname{appendix}{Appendix}{Appendices}
\Crefname{appendix}{Appendix}{Appendices}

\DeclareMathOperator{\sgn}{sgn}

\newcommand{\bra}[1]{\langle #1|}
\newcommand{\ket}[1]{|#1\rangle}
\newcommand{\braket}[1]{\langle #1 \rangle}

\newcommand{\e}{e}

\newcommand{\dd}{d}

\newcommand{\half}{\frac{1}{2}}

\newcommand{\hc}{\text{H.c.}}

\newcommand{\dg}{^\dagger}

\def\*#1{\mathbf{#1}}

\definecolor{darkgreen}{rgb}{0,0.6,0}

\makeatletter
\AtBeginDocument{\let\LS@rot\@undefined}
\makeatother

\begin{document}

\title{
Majorana Qubit Readout Using Longitudinal Qubit-Resonator Interaction
}

\author{Arne L. Grimsmo}\email{arne.grimsmo@sydney.edu.au}
\affiliation{Centre for Engineered Quantum Systems, School of Physics, The University of Sydney, Sydney, Australia}
\author{Thomas B. Smith}
\affiliation{Centre for Engineered Quantum Systems, School of Physics, The University of Sydney, Sydney, Australia}

\date{\today}

\begin{abstract}
We propose a quantum non-demolition Majorana qubit readout protocol based on parametric modulation of a longitudinal interaction between a pair of Majorana bound states and a resonator. The interaction can be modulated through microwave frequency gate-voltage or flux control of a tunable tunnel barrier. The qubit-resonator coupling is quantum non-demolition to exponential accuracy, a property inherited from the topological nature of the Majorana zero modes. 
The same mechanism as used for single-qubit readout can also be extended to perform multi-qubit measurements and be used to enact long range entangling gates.
\end{abstract}

\maketitle

\section{Introduction}
When quantum information is stored in Majorana zero modes (MZMs) in topological superconductors, appreciable overlap of two Majorana wavefunctions leads to a splitting of the ground space degeneracy. Quantum information can then in principle be read out by measuring the resulting localized charge. Promising progress has been made towards realizing MZMs in epitaxial semiconductor-superconductor nanowire hybrids~\cite{Mourik2012,Deng2016,Albrecht2016} and in 2D hybrids with litographically defined 1D channels~\cite{Nichele2017,Zhang2018,Suominen2017}. However, many questions remain about how to best control and measure these systems. To fully harness the long lifetimes expected for quantum information stored in topological systems, it is crucial that the logical operations are fast and do not introduce new errors that break the inherent protection from noise.

Measurement-only topological quantum computing is an attractive approach to computing with MZMs, because it removes the need to physically transport anyons~\cite{Bonderson08}. The price paid is that a single physical braid is replaced by a series of measurements, where the total number of measurements needed is probabilistic~\cite{Plugge2017,Karzig2017}. Qubit readout is known to be rather slow compared to unitary gates in more developed solid-state qubit architectures~\cite{Barends2014,Walter17,west2018gate}. This raises a question of whether readout of MZMs can be made sufficiently fast and high fidelity to allow gates with better performance than conventional qubits.

We propose an approach to readout of MZMs based on parametric modulation of a longitudinal interaction between a pair of Majorana modes and a single mode of a readout resonator. MZMs localized at the ends of two topological superconductors overlap across a non-proximitized semiconducting barrier, forming a gate-tunable valve~\cite{Aasen2016}.
A schematic of such a setup is shown in~\cref{fig:setup}.
Interaction with the electromagnetic field is in turn introduced by capacitive coupling to a resonator.
The non-local nature of the Majorana modes implies that the coupling to the resonator voltage is proportional to the overlap of a pair of Majorana wavefunctions, giving an interaction of the form $ \hat H_\text{int}  = -e \eta \hat V_r \hat\gamma_i\hat\gamma_j$, where $\hat V_r$ is the voltage bias of the resonator, $\hat\gamma_{i,j}$ are Majorana operators which satisfy $\hat \gamma_i = \hat \gamma_i\dg$ and $\{\hat \gamma_i, \hat \gamma_j\} = 2\delta_{ij}$, and $\eta$ is proportional to the Majorana wavefunction overlap.

In itself, $\hat H_\text{int}$ is not directly useful for readout because there is no energy exchange between the resonator and the Majorana subsystem, and the interaction leads to a negligible response of the resonator. To enact a measurement we therefore propose to modulate the Majorana wavefunction overlap, $\eta \to \eta(t)$. In practice, this can be done through gate voltages or an external flux controlling the energy cost of electron tunneling, as we will show.
If $\eta(t)$ is modulated at the resonator frequency, $\hat H_\text{int}$ takes the form of an on-resonance ac voltage drive of the resonator, with a Majorana state dependent phase. As a consequence, the resonator field is displaced in one of two diametrically opposite directions in phase space, depending on the eigenvalue $\pm 1$ of $i\hat \gamma_i\hat\gamma_j$, leading to a large response for the resonator

The scheme proposed here is based on a readout protocol introduced in Ref.~\cite{Didier2015} which relies on longitudinal (as opposed to transverse) qubit-resonator interaction. While this approach was originally introduced in the context of transmon qubits, longitudinal coupling is the \emph{natural} form for the light-matter interaction with bound Majorana modes~\cite{Dartiailh2017}.
The quantum non-demolition (QND) nature of the coupling is in a certain sense protected by the fractional and non-local nature of the MZMs. As the coupling is only present for Majorana modes with non-negligible overlap, one can controllably choose to read out a single pair $i\hat \gamma_i\hat \gamma_j$, while coupling to all other MZMs is exponentially suppressed. The observable $i\hat \gamma_i \hat \gamma_j$ is thus a constant of motion, leading to a measurement that is QND, up to exponentially small corrections.
We refer to this manifestation of the topological nature of Majorana bound modes in a measurement, first discussed in Ref.~\cite{Plugge2017}, as topological QND (TQND) measurement.

We also show how the readout protocol can be extended to multi-qubit measurements, and how the same mechanism proposed for the readout protocol can be used to enact long-range entangling gates between Majorana qubits. The proposed two-qubit gate is non-topological, but is nevertheless compelling as an ingredient in a topological quantum computing platform. 
It can be used to generate entangled qubit pairs used in two-qubit gate teleportation~\cite{Bravyi2006}, and the possibility of generating long-range entanglement can facilitate a more modular approach to topological quantum computing~\cite{Nickerson2013}.

\section{Longitudinal Majorana-resonator interaction}
\begin{figure}
\centering
\includegraphics{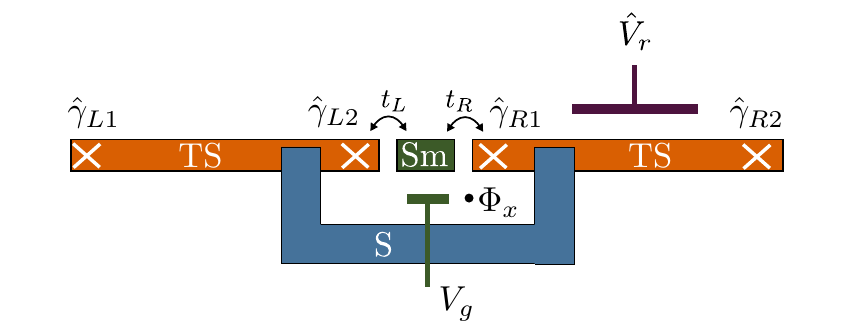}
\caption{\label{fig:setup}Schematic Majorana qubit setup. Two topological superconducting wires (TS) host four Majorna bound states, located at the respective wire ends (crosses). A dot is formed in a non-proximitized semiconducting segment (Sm), acting as a tunable barrier between the two wires. A conventional superconductor (S) shunts the two wires such that they behave as a single superconducting island with a uniform superconducting phase. The superconducting island furthermore couples capacitively to the voltage of a resonator ($\hat V_r$). The overlap of the Majorana wavefunctions can be controlled through gate-voltages ($V_g$), or through an external flux threading the superconducting loop ($\Phi_x$). 
A geometry with two parallel wires as in Ref.~\cite{Plugge2017} can also be used.
}
\end{figure}

We consider physical realizations of Majorana qubits where pairs of Majorana bound states are localized at the ends of a quasi one-dimensional electronic system, or quantum wire. Light-matter interaction arises due to capacitive coupling to the electric field of a resonator.
To be specific, consider the schematic setup illustrated in~\cref{fig:setup}. Two topolological superconductors are tunnel coupled to a common semiconducting (Sm) barrier,
effectively forming a topological superconductor-semiconductor-superconductor (TS-Sm-TS) junction.
We assume that the barrier behaves as a quantum dot with a discrete set of orbitals. This can be a naturally formed dot in a non-proximitized segment of a superconductor-semiconductor hybrid where a single gate voltage can be used to control the dot energy levels~\cite{Deng2016,Deng2018}. Alternatively, additional gates can be introduced to directly control the tunnel coupling between the dot and the wires, as in the proposals in Refs.~\cite{Plugge2017,Karzig2017}.
We focus on a situation where the dot levels are off resonant from the Majorana modes such that the dot effectively acts as a tunable barrier.

We moreover assume that each wire is in contact with a common non-topological superconductor (denotes S in the figure), such that
we can treat the two wires and the conventional superconductor as a single superconducting island~\cite{Plugge2017}.
Finally, both the superconducting island and the semiconducting barrier can be capacitively coupled to the the electric field of a resonator (in the figure, only capacitive coupling to the superconducting island is indicated).

As discussed in more detail in~\cref{app:interaction}, we can write a low-energy Hamiltonian for the Majorana qubit (TS-Sm-TS) system excluding the resonator field as
\begin{equation}\label{eq:Hq}
    \hat H_q =
    \sum_{\alpha=L,R} E_\alpha i \hat\gamma_{\alpha 1} \hat \gamma_{\alpha 2}
    + \hat H_C
    + \hat H_B
    + \hat H_T,
\end{equation}
where the index $\alpha$ labels the two wires.
The bare Majorana fermion splittings of each wire $E_{\alpha} \propto \e^{-L_\alpha/\xi_\alpha}$
is assumed to be exponentially small, where $L_\alpha$ is the wire length and $\xi_\alpha$ a characteristic coherence length~\cite{Knapp2018}. We will consider only the ideal long wire regime and therefore take $E_{\alpha}\to 0$ from here on.
The term $\hat H_{C} = E_{C} (\hat N - n_g)^2$
describes the charging energy of the superconducting island, with $\hat N$ the island's electron occupancy and $n_g$ an offset charge.
The two last terms in~\cref{eq:Hq} describe the barrier between the left and right wires, which we model by Hamiltonians
\begin{equation}
    \hat H_\text{B} = \sum_j h_j \hat b_j\dg \hat b_j + U\left( \sum_j \hat b_j\dg \hat b_j - n_b\right)^2
\end{equation}
and~\cite{Fu2010}
\begin{equation}\label{eq:Htunn}
    \hat H_T = \sum_j \frac{i t_{Lj}}{2} \e^{\frac{i(\hat\varphi + \varphi_x)}{2}}\hat \gamma_{L2}\hat b_j - \frac{t_{Rj}}{2} \e^{\frac{i\hat\varphi}{2}} \hat \gamma_{R1} \hat b_j
    + \hc
\end{equation}
Here $h_j$ are bare energies for orbitals localized in the semiconducting barrier in the uncoupled limit ($t_{\alpha j}\to 0$), and $\hat b_j\dg$ the corresponding fermionic creation operators. $U$ is the dot charging energy and $n_b$ a gate-tunable offset charge, while
$t_{\alpha j}\ge 0$ are tunneling amplitudes proportional to the overlap of the corresponding Majorana mode functions and the dot-orbitals (see~\cref{app:interaction}). 
The superconducting phase $\hat \varphi$ appearing in the tunneling Hamiltonian and the island electron number $\hat N$ are canonical conjugate variables satisfying $[\hat N,\e^{i\hat \varphi/2}] = \e^{i\hat \varphi/2}$.
We also include an external flux $\varphi_x = 2\pi \Phi_x/\Phi_0$, with $\Phi_0=h/2e$ the magnetic flux quantum, threading the loop formed by the TS-Sm-TS junction and the bulk superconductor.

Coupling to the electromagnetic field is introduced through a total Hamiltonian $\hat H = \hat H_{q} + \hat H_r + \hat H_\text{int}$ where $\hat H_r = \hbar\omega_r \hat a\dg \hat a$ and
\begin{equation}\label{eq:Hint}
\hat H_\text{int} = \sum_\alpha i \hbar \lambda_{C}\hat N(\hat a\dg - \hat a)
+ \sum_j i \hbar \lambda_j \hat b_j\dg \hat b_j(\hat a\dg - \hat a).
\end{equation}
Here $\hat a\dg$ creates a resonator photon with energy $\hbar\omega_r$, and $\lambda_C$ ($\lambda_j$) is a coupling constant describing capacitive coupling of the resonator to the island charge (the semiconducting barrier).

In the proposed readout protocol, the barrier occupation energies are gradually lowered (alternatively, the tunnel couplings $t_{\alpha j}$ are gradually turned on) such that the initial (near) zero energy logical qubit eigenstates evolve into hybridized states partially localized in the semiconducting barrier. The logical states then become split in energy and couple to the resonator field. The key physics behind the effective Majorana-resonator coupling can be exposed by diagonalizing $\hat H_q$ and modeling the semiconducting barrier by a single orbital $j=0$ for simplicity,
$\hat H_B = \varepsilon_0 \hat b_0\dg \hat b_0$ with $\varepsilon_0 = h_0 + U(1-2n_g)$.
Since $\hat H_T$ only couples states $\ket{N=n,n_0=0}$ and $\ket{N=n-1,n_0=1}$, where $N$ is an island charge eigenvalue and $n_0$ the barrier occupancy, we can diagonalize $\hat H_{q}$ block by block. 
We here restrict our focus to $n_g \simeq 0$ and large charging $E_C$ and barrier energy $\varepsilon_0$, such that relevant subspace is spanned by $\{\ket{N=0,n_0=0},\ket{N=-1,n_0=1}\}$. The general case is given in~\cref{app:diagonalization}.
Furthermore, treating the resonator-interaction $\hat H_\text{int}$ as a perturbation, we have a Hamiltonian for the low-energy subspace
\begin{equation}\label{eq:Hpert}
    \begin{aligned}
    \hat H \simeq{}& 
    \hbar \omega_r \hat a\dg \hat a
    + \frac{\hbar \omega_q}{2} \hat \sigma_z
    + i\hbar g_z (\hat \sigma_z+1) (\hat a\dg - \hat a),
    \end{aligned}
\end{equation}
where we have defined $\hat\sigma_z = i\hat \gamma_{L2}\hat\gamma_{R1}$, 
and to first order in the resonator couplings we have $\hbar \omega_q = \half\left[f_+(\varphi_x) - f_-(\varphi_x)\right]$
and $g_z = -\frac{\lambda_C-\lambda_0}{4}\frac{\partial \hbar\omega_q}{\partial \delta}$,
with $\delta = \varepsilon_0+E_C(1+2n_g)$ and
$f_\pm(\varphi_x) = \sqrt{\delta^2 + t_L^2 + t_R^2 \pm 2t_Lt_R\cos\left(\frac{\varphi_x}{2}\right)}$.
In the limit of small $t_L, t_R \ll \delta$ we have the simplified expressions
$\hbar\omega_q \simeq t_Lt_R\cos\left(\varphi_x/2\right)/\delta$ and
$g_z \simeq \frac{\lambda_C-\lambda_0}{4}t_Lt_R\cos\left(\varphi_x/2\right)/\delta^2$.

We emphasize that it is the \emph{difference} in the bare coupling strength $\delta_\lambda \equiv \lambda_0-\lambda_C$ that appears in the low-energy effective Majorana-resonator coupling $g_z$. In other words, the resonator is sensitive to the difference in voltage bias depending on if a charge is localized in the superconductor or in the semiconductor. The coupling differential $|\delta_\lambda|$ can be maximized by appropriately fabricating coupling capacitances.

Note that~\cref{eq:Hpert} assumes that $\delta_\lambda$ is sufficiently small and the energy $\delta$ of moving an electron from the island into the barrier is sufficiently large such that resonator induced coupling to excited states in the barrier can be neglected. Higher order terms in a perturbative expansion in $\delta_\lambda$ might change the dynamics of the resonator, but will nevertheless commute with $\hat\sigma_z$ (see~\cref{app:diagonalization} for the full expression).

\section{Parametric qubit readout}

\begin{figure}
\centering
\includegraphics{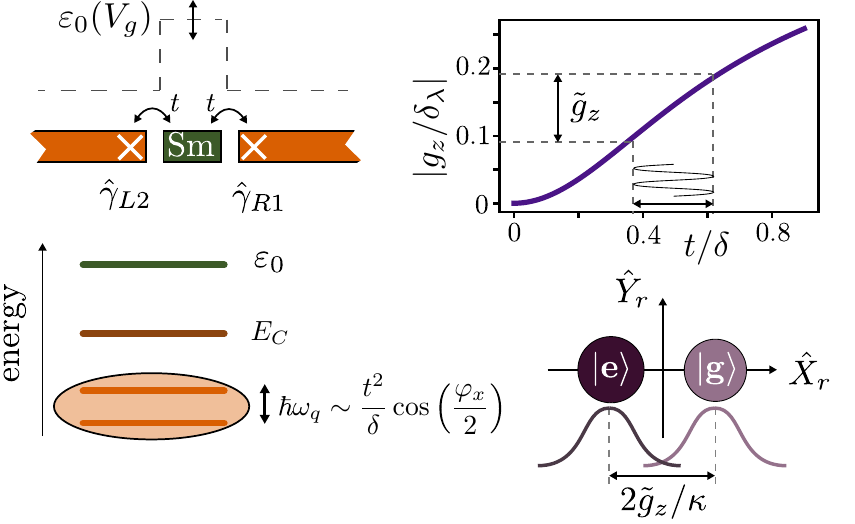}
\caption{\label{fig:concept}Longitudinal readout concept: A semiconducting segment (Sm) forms a dot acting as a tunable barrier between the topological quantum wires. 
By modulating either the tunneling amplitudes, barrier excitation energy, or an external flux through the qubit loop, the longitudinal coupling can be modulated with an amplitude $\tilde g_z$, leading to a qubit state dependent displacement of the resonator with steady state magnitude $\pm \tilde g_z/\kappa$.}
\end{figure}

The Majorana-resonator coupling in~\cref{eq:Hpert} is of a longitudinal nature.
Longitudinal coupling here refers to a qubit-resonator Hamiltonian interaction of the form $\hat H_z = ig_z \hat\sigma_z (\hat a\dg - \hat a)$ that is proportional to the qubit Hamiltonian $\hat H_q \propto \hat \sigma_z$, while transversal coupling in contrast refers to an interaction $\hat H_x = ig_x \hat \sigma_x(\hat a\dg - \hat a)$ orthogonal to the qubit Hamiltonian.
At first glance, the longitudinal nature of the qubit-resonator interaction might seem like a disadvantage, because there is no energy exchange between the two systems (note that the coupling terms are fast rotating in the interaction frame, and would thus average out on a short time-scale). 
In a proposal introduced in Ref.~\cite{Ohm2015} this issue was overcome by working with relatively short wires, introducing terms of the form $E_\alpha i\hat\gamma_{\alpha 1}\hat\gamma_{\alpha 2}$ to the Hamiltonian. However, this leads to a readout that is not truly QND, and breaks the topological protection of the qubit.
We instead propose to work in the long wire limit with purely longitudinal coupling (up to exponentially small corrections in the wire length), and introduce a simple cure to bridge the energy gap between the logical qubit states and the resonator~\cite{Didier2015}: By parametrically modulating the longitudinal coupling strength at the resonator frequency the longitudinal interaction takes the form of a resonant drive of the resonator, with a qubit-state dependent phase, leading to a large response for the resonator. Note that in contrast to readout based on transversal coupling, the standard approach for superconducting qubits~\cite{Blais2004}, a longitudinal interaction leads to a fully QND readout. The readout protocol is illustrated conceptually in~\cref{fig:concept}.

\begin{figure}
\includegraphics{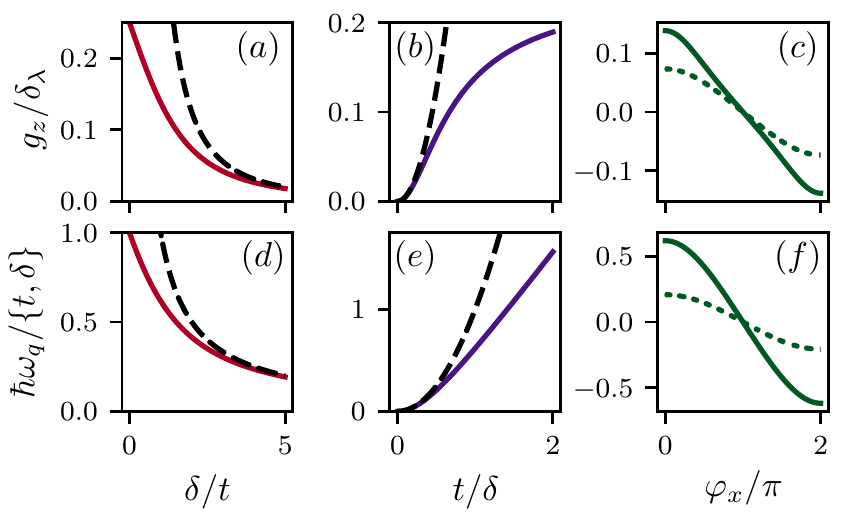}
\caption{\label{fig:spectrum}
$(a)$--$(c)$: Parametric dependence of longitudinal coupling on $\delta/t$ $(a)$ and $t/\delta$ $(b)$ for $\varphi_x=0$, and on $\varphi_x$ $(c)$ for $t/\delta=1$ (solid line) and $t/\delta=0.5$ (dotted line).
$(d)$--$(e)$: Qubit splitting $\hbar\omega_q/t$ $(d)$, $\hbar\omega_q/\delta$ $(e,f)$ for the same parameters as in the top row.
The dashed lines in $(a)$, $(b)$, $(d)$ and $(e)$ are approximations for small $t/\delta$. The parameter $\delta_\lambda=\lambda_0-\lambda_C$ is the difference in resonator-coupling strength for the semiconducting barrier and the superconducting island.
}
\end{figure}

The barrier energies $\varepsilon_j$ and/or tunnel couplings $t_{\alpha j}$ can be controlled via gate voltages, giving us a mechanism for modulating the Majorana-resonator coupling, $\omega_q \to \omega_q(V_g)$, $g_z \to g_z(V_g)$. Effectively, this amounts to modulating the Majorana wavefunction overlap across the semiconducting barrier.
Another attractive option is to modulate the external flux $\varphi_x(t) = \bar\varphi_x + \tilde\varphi_x(t)$, since for small tunneling we have shown that $g_z \sim \cos(\varphi_x/2)$.
In~\cref{fig:spectrum} we show the parametric dependence of $g_z$ and $\omega_q$ on $\delta$, $t=t_L=t_R$ and external flux $\varphi_x$, for a single barrier-orbital $j=0$ as before. Notably, the coupling $g_z$ can be a large fraction of the bare coupling differential $|\delta_\lambda| = |\lambda_0-\lambda_C|$.

Following Ref.~\cite{Didier2015}, we propose to modulate the effective coupling strength $g_z(t) = \bar g_z + \tilde g_z\cos(\omega_r t)$ at the resonator frequency. In the interaction picture and dropping fast rotating terms we then have $\tilde H_\text{ideal} = \half \tilde g_z \hat \sigma_z(\hat a\dg - \hat a)$.~\footnote{The analytical expressions for $g_z$ and $\omega_q$ displayed in~\cref{fig:spectrum} were derived assuming a time-independent Hamiltonian. Modulation of the system parameters will lead to corrections to these parameters.}.
The dynamics under this Hamiltonian is exactly solvable: In the long-time limit the resonator is displaced to one of two coherent states $\ket{\pm \alpha}$ depending on the qubit state, with $\alpha = \tilde g_z/\kappa$ where $\kappa$ is the resonator decay rate~\cite{Didier2015}.

\begin{figure}
\centering
\includegraphics{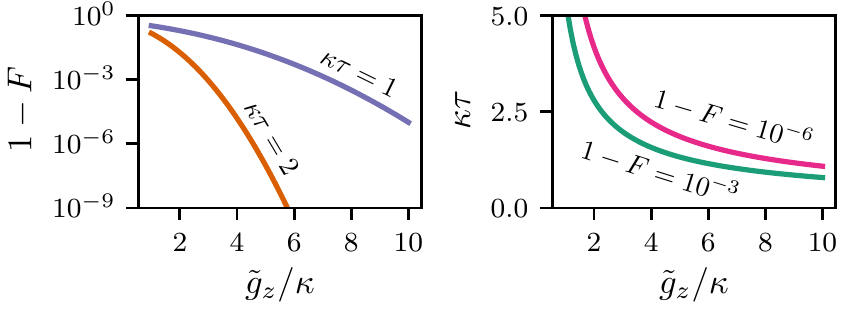}
\caption{\label{fig:fidelity}Measurement infidelity for ideal longitudinal readout with interaction $\hat H_\text{ideal}$. $(a)$ Infidelity as a function of modulation amplitude for two different measurement times $\kappa\tau=1,2$. $(b)$ Measurement time needed to reach infidelities $1-F=10^{-3}$ and $1-F=10^{-6}$ as a function of modulation amplitude.}
\end{figure}

As was demonstrated in Ref.~\cite{Didier2015} parametric modulation of longitudinal coupling can lead to extremely fast, QND readout.
We illustrate this with the idealized model $\tilde H_\text{ideal}$ in~\cref{fig:fidelity}.
Panel~$(a)$ shows the readout infidelity $1-F$ as a function of $\tilde g_z/\kappa$ for two different measurement times $\kappa\tau = 1, 2$, and panel~$(b)$ shows the measurement time and modulation amplitude needed to reach infidelities $1-F=10^{-3}$ and $1-F=10^{-6}$. 
Details on the calculation of $F$ are given in~\cref{app:fidelity}.
One of the remarkable properties of longitudinal qubit readout is the fast rate at which qubit information is attained~\cite{Didier2015}, as these results show.
For example, for a readout rate of $\kappa/(2\pi) = 1$ MHz and a coupling modulation amplitude of $\tilde g_z/(2\pi) = 5$ MHz, an infidelity of $10^{-6}$ can be reached in about 300 ns for the ideal model. 

We emphasize that there are two ingredients necessary for realizing fast, high-fidelity longitudinal readout: Both a longitudinal qubit-resonator interaction, and fast control of the interaction strength to allow modulation at the resonator frequency. The combination of a highly robust longitudinal interaction protected by the non-local nature of MZMs, and the natural parametric control over the interaction in Majorana qubit systems through gate-voltages or external flux, makes this qubit architecture especially attractive for realizing longitudinal readout.

Realistically we expect reduction in the measurement fidelity compared to the idealized results presented in~\cref{fig:fidelity}.
We can distinguish between two qualitatively different types of noise sources: Those that preserve the QND nature of the interaction and those that break it.
For example, charge and flux noise lead to fluctuations in the coupling strength $g_z$, which might reduce the signal-to-noise ratio and thus the measurement fidelity for a fixed readout time. However, this does not change the longitudinal form of the interaction Hamiltonian. Similarly, higher-order terms neglected in~\cref{eq:Hpert} might alter the dynamics of the resonator, \emph{e.g.} cross- and self-Kerr terms of the form $\sim \hat \sigma_z \hat a\dg \hat a$ and $\sim \hat a^{\dagger 2} \hat a^2$, but any higher order term will nevertheless commute with $\hat \sigma_z$ as long as transitions out of the low-energy logical subspace can be neglected. 
Even more remarkably, the same argument holds for coupling to the electromagnetic field beyond the readout resonator as well (see~\cref{app:env}). The electromagnetic environment thus only causes dephasing noise in the measurement basis, leading to the notion of a TQND measurement.
In contrast to the mentioned noise channels are noise channels that cause state transitions in the measurement basis. Two likely causes of this are finite wire lengths~\cite{Knapp2018} and quasi-particle poisoning events~\cite{Rainis2012,Knapp2018modelingnoiseerror}.
These errors explicitly break the assumptions behind the topological protection of the qubit and cause faulty measurement outcomes which can not be overcome by measuring for longer times.

\section{Long-range interactions}

The same mechanism as used for readout, \emph{i.e.}, parametric modulation of the longitudinal qubit-resonator interaction, can also be used to enact long-range two-qubit gates. Consider a two-qubit setup, where for each qubit a pair of Majorana operators are tunnel coupled a semiconducting barrier, as before. Each qubit island is in turn coupled capacitively to a common resonator mode. Based on the results leading up to~\cref{eq:Hpert} we expect this setup to be well described by a Hamiltonian
\begin{equation}\label{eq:Htwoqubit}
    \begin{aligned}
    \hat H ={}& \frac{\hbar\omega_{q1}(t)}{2}i\hat\gamma_1\hat\gamma_2 + \frac{\hbar\omega_{q2}(t)}{2}i\hat\gamma_5\hat\gamma_6 + \hbar\omega_r\hat a\dg \hat a \\
    +& i\hbar g_1(t)i\hat\gamma_1\hat\gamma_2 (\hat a\dg - \hat a)
    + i\hbar g_2(t)i\hat\gamma_5\hat\gamma_6 (\hat a\dg - \hat a).
    \end{aligned}
\end{equation}
Here, qubit one is defined in terms of Majorana operators $\hat\gamma_1,\dots,\hat\gamma_4$ and qubit two in terms of $\hat \gamma_5,\dots,\hat\gamma_8$.
Modulating the coupling parameters at the resonator frequency leads to a readout of the two observables $i\hat \gamma_1\hat\gamma_2$ and $i\hat \gamma_5\hat\gamma_6$, as before. In contrast, by choosing a modulation frequency $g_i(t) = \bar g_i + \tilde g_i\cos(\omega_mt)$ which is far off-resonant $|\omega_r-\omega_m| \gg \kappa$, \cref{eq:Htwoqubit} leads to an effective qubit-qubit interaction. As shown in Ref.~\cite{Royer2017}, an \emph{exact} unitary transformation maps the Hamiltonian~\cref{eq:Htwoqubit} onto $\hat H' = \hbar J \hat \gamma_1\hat \gamma_2 \hat \gamma_5 \hat \gamma_6 + \hbar\omega_r \hat a\dg \hat a$, where $J \simeq \tilde g_1\tilde g_2/(\omega_m-\omega_r)$. By modulating for a period of time $t_g = \pi/4|J|$ this gives a two-qubit entangling gate. In an encoding where $i\hat\gamma_1\hat\gamma_2 = \hat\sigma_{z1}$ and $i\hat\gamma_5\hat\gamma_6 = \hat\sigma_{z2}$ the gate is, up to single qubit unitaries, equivalent to a controlled-$Z$ gate~\cite{Royer2017}. If the tunnel couplings are tunable such that one can selectively couple to different pairs of Majorana operators on each island, similar to the proposals in Ref.~\cite{Karzig2017}, it is furthermore possible to enact effective $\bar P_i \otimes \bar P_j$ interactions where $\bar P_i, \bar P_j$ are any of $\hat \sigma_x,\hat \sigma_y, \hat \sigma_z$.

We emphasize that the proposed two-qubit gate is not topologically protected: The qubit degeneracy is lifted during the gate. Moreover, as resonator photons leak out at rate $\kappa$, this leads to resonator-induced qubit dephasing~\cite{Royer2017}.

\section{Two-qubit parity measurements}

Single-qubit readout is in itself an important primitive in any Majorana based quantum computation scheme. To enact the full set of Clifford gates in a measurement-only scheme, however, it is furthermore necessary to measure four-MZM terms of the type $\hat \gamma_1\hat \gamma_2\hat \gamma_3 \hat \gamma_4$~\cite{Karzig2017}.

\begin{figure}
\includegraphics{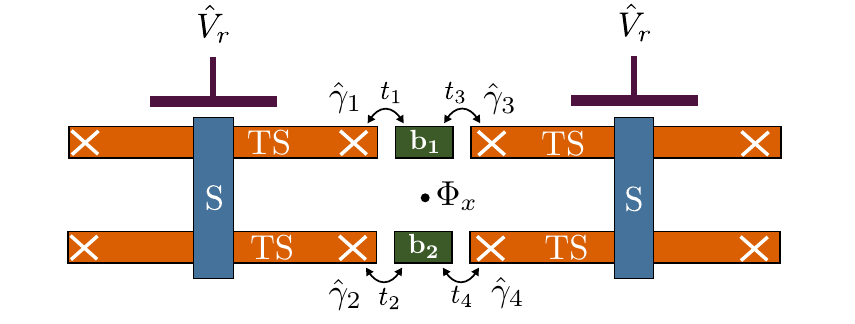}
\caption{\label{fig:4mzm}Schematic of two Majorana box qubits interfaced by two separate semiconducting barriers ($b_1, b_2$), and capacitively coupled to a common readout resonator voltage $\hat V_r$. In the figure capacitive coupling to each superconducting island is indicated, but it is sufficient that one of the two qubits couple to the resonator.}
\end{figure}

An example setup is shown in~\cref{fig:4mzm}, where two Majorana box qubits (left and right) are interfaced via two distinct semiconducting barriers (denoted by $b_1$ and $b_2$), as proposed in Ref.~\cite{Plugge2017}. When the energy cost of moving a charge off any of the two superconducting islands is large and the relevant barrier orbitals are unoccupied, this setup leads to an effective low-energy interaction $\sim \hat \gamma_1\hat\gamma_2\hat\gamma_3\hat\gamma_4$, which is in a certain sense protected. As explained in Ref.~\cite{Karzig2017} this can be seen by considering the tunneling paths for a fermion to start and end up on the same island: Either it can travel partway around the loop and backtrack, which leads to an operator proportional to the identity in perturbation theory (\emph{e.g.}, $\hat \gamma_1\hat \gamma_3\hat \gamma_3\hat \gamma_1 = 1$), or it can make the full loop, leading to the desired combination $\hat \gamma_1\hat\gamma_2\hat\gamma_3\hat\gamma_4$. Note, however, that this argument does not hold if an electron is shared between two near-resonant states for the two Sm-barriers, since the charge can tunnel from one barrier to the other leading to terms proportional to $\hat\gamma_1\hat\gamma_2$ and $\hat \gamma_3\hat\gamma_4$ in the effective low-energy Hamiltonian. We therefore focus on a setup where the relevant barrier orbitals are intialized to be empty.

To introduce a readout mechanism, we assume that the superconducting islands and the barrier orbitals are capacitively coupled to a single readout resonator, as before. Considering only one orbital for each barrier, with fermion annihilation operators $\hat b_1$ and $\hat b_2$, respectively, a Hamiltonian describing the setup illustrated in~\cref{fig:4mzm} is
\begin{equation}
    \begin{aligned}
    \hat H ={}& 
    \hat H_0 + \hat H_r
    + \hat H_T + \hat H_\text{int}
    \end{aligned}
\end{equation}
where
\begin{align}
\hat H_0 ={}& E_{L} \hat N_L^2 + E_{R} \hat N_R^2 + \varepsilon_1 \hat b_1\dg \hat b_1 + \varepsilon_2 \hat b_2\dg \hat b_2,\\
\hat H_r ={}& \hbar\omega_r \hat a\dg \hat a,
\end{align}
are bare Hamiltonians for the Majorana qubit and resonator subsystems, respectively,
\begin{equation}
    \begin{aligned}
    \hat H_T ={}& \e^{i\hat\varphi_L/2}\left(t_1 \hat \gamma_1 \hat b_1 + \hat t_2 \hat \gamma_2 \hat b_2 \right) \\
    & +\e^{i\hat \varphi_R/2}\left(t_3 \hat\gamma_3 \hat b_1 + t_4 \hat\gamma_4 \hat b_2 \right) + \hc
    \end{aligned}
\end{equation}
is a tunneling Hamiltonian, and
\begin{equation}
    \begin{aligned}
    \hat H_\text{int} ={}& \left(\sum_{\nu=L,R}\lambda_\nu \hat N_\nu + \sum_{i=1,2} \lambda_{i}\hat b_i\dg \hat b_i \right)i(\hat a\dg - \hat a)
    \end{aligned}
\end{equation}
describes capacitive coupling to the resonator. In these expressions, $\varepsilon_i$ are orbital occupation energies, $E_\nu$ are charging energies for the left ($\nu=L$) and right ($\nu=R$) island, and $\hat N_\nu$ and $\hat \varphi_\nu$ the corresponding charge and phase operators, satisfying $[\hat N_\nu,\e^{i\hat\varphi_\mu/2}]=\delta_{\nu\mu}\e^{i\hat\varphi_\mu/2}$. We have neglected any offset charge, for simplicity, and any external flux $\varphi_x = 2\pi \Phi_x/\Phi_0$ through the qubit loop can be absorbed into the tunnel couplings $t_i$. Finally, $\lambda_x$ for $x=L,R,1,2$ describes capacitive coupling of the different parts to the readout resonator.

To illustrate the principle behind longitudinal coupling for this setup, we treat $\hat H_T + \hat H_\text{int}$ as a perturbation,
and perform a Schrieffer-Wolff transformation with respect to the qubit subsystem $\hat H_0$ only. For large charging and barrier occupation energies, the relevant low-energy state is $\ket{\Omega}\equiv \ket{N_L=0,N_R=0,n_1=0,n_2=0}$, where $N_\nu$ refers to the charge on island $\nu$ and $n_i$ the occupation of the $i$th barrier. Following Ref.~\cite{bravyi2011schrieffer} we define projection operators onto the ground state $\ket \Omega$ and excited states of $\hat H_0$ and find an effective low-energy Hamiltonian $\hat H_\text{eff} = \braket{\Omega|\hat U\dg \hat H \hat U|\Omega} = \hat H_0 + \hat H_1 + \hat H_2 + \dots$ To fifth order in Schrieffer-Wolff, and neglecting fast rotating terms for the resonator, we find
\begin{equation}\label{eq:4zmzeff}
    \begin{aligned}
    \hat H_\text{eff} ={}& \hat H_r - (A+A^*) \hat \gamma_1\hat\gamma_2\hat\gamma_3\hat\gamma_4\\
    +& (B+B^*) \hat \gamma_1\hat\gamma_2\hat\gamma_3\hat\gamma_4 i(\hat a\dg - \hat a),
    \end{aligned}
\end{equation}
where we have dropped a frequency shift of the resonator, and the coefficients are given by
\begin{align}
    &A = \frac{t_1 t_3^* t_4^* t_2}{\Delta_1\Delta_2\Delta_5} + \frac{t_1 t_3^*  t_4^* t_2}{\Delta_3\Delta_4\Delta_6},\\
    &\begin{aligned}
        B{=}& \frac{t_1t_3^* t_4 t_2^* }{\Delta_1\Delta_2\Delta_5}\left(\frac{\lambda_1-\lambda_L}{\Delta_1} + \frac{\lambda_2-\lambda_L}{\Delta_2} + \frac{\lambda_R - \lambda_L}{\Delta_5} \right) \\
        +&{}\frac{t_1t_3^* t_4 t_2^* }{\Delta_3\Delta_4\Delta_6}\left(\frac{\lambda_1-\lambda_R}{\Delta_3} + \frac{\lambda_2-\lambda_R}{\Delta_4} + \frac{\lambda_L-\lambda_R}{\Delta_6} \right),
    \end{aligned}
\end{align}
with $\Delta_1 = E_L + \varepsilon_1$, $\Delta_2 = E_L + \varepsilon_2$, $\Delta_3 = E_R + \varepsilon_1$, $\Delta_4 = E_R + \varepsilon_2$, $\Delta_5 = \Delta_6 = E_L + E_R$.

Again, this result suggests a way to read out the observable $\hat\gamma_1\hat\gamma_2\hat\gamma_3\hat\gamma_4$ by modulating the longitudinal coupling strength. This can be done by varying the tunnel rates, or since $t_1t_3^*t_4^*t_2 \sim \cos(\varphi_x/2)$, the external flux through the qubit loop. We emphasize that the basic idea of multi-MZM measurements protected by an energetic constraint is not restricted to longitudinal readout using parametric modulation. In particular, dispersive coupling terms will occur at higher order in perturbation theory. There is also, clearly, a cost to measuring higher weight MZM operators. The idea can in principle be extended to measuring any even-weight string of Majorana operators, but at the cost of a coupling to the resonator occurring at even higher order in perturbation theory.

\section{Conclusions}

We have introduced a readout protocol for Majorana qubits based on parametric modulation of a longitudinal Majorana-resonator interaction. Under modest assumptions about the magnitude of the coupling modulation, our results suggest that high-fidelity readout is possible in short measurement times. 
Fast and high-fidelity readout is equally important to long coherence times in measurement-only topological quantum computing, as measurement time as a fraction of coherence time ultimately bounds the overall fidelity of logical braids.
Moreover, the same mechanism as used for readout can be used to generate long-range entanglement between Majorana qubits.
The hardware requirements for coupling to resonators and the necessary parametric control is similar to other solid-state qubit architectures, including superconducting qubits~\cite{Barends2014,corcoles2015demonstration},
 thus making parametric modulation of longitudinal coupling an attractive ingredient in a topological quantum computing architecture.

\begin{acknowledgments}
We thank Stephen Bartlett, Maja Cassidy, Andrew Doherty, Karl Petersson and David Reilly for helpful discussions.
This work is supported by the Australian Research Council (ARC) via Centre of Excellence in Engineered Quantum Systems (EQUS) Project No. CE170100009 and DECRA No. DE190100380.
\end{acknowledgments}

\appendix

\section{\label{app:interaction}Tunable interaction between a pair of Majorana modes and a resonator}

\subsection{A single superconducting island hosting two quantum wires}

We start by considering two quantum wires, each hosting a pair of bound Majorana modes, on a single superconducting island with charging energy $E_C$.
We for now ignore the tunnel coupling to the semiconducting region (labelled ``Sm'' in the~\cref{fig:setup}). We return to the tunnel coupling in the next section. The superconducting island hosting the two quantum wires can be described by a Hamiltonian
\begin{equation}\label{eq:Halpha}
    \hat H_\text{island} = E_{C}(\hat N + \hat n_e - n_g)^2 + \sum_{\alpha=L,R} \hat H_{e,\alpha},
\end{equation}
where the first term is the charging energy of the superconducting island, with $\hat N$ equal to two times the number of Cooper pairs and 
\begin{equation}
    \hat n_e = \sum_{\alpha=L,R} \sum_{\sigma=\uparrow,\downarrow} \int \dd^3 r \hat \psi_{\alpha\sigma}\dg(\*r)\psi_{\alpha\sigma}(\*r),
\end{equation}
counts any unpaired, near-zero energy electrons~\cite{Hell16}.
Here $\hat\psi_{\alpha\sigma}(\*r)$ is an electron field for the quantum wire labeled $\alpha \in \{L,R\}$ (with spin $\sigma=\uparrow,\downarrow$) satisfying $\{\hat\psi_{\alpha\sigma}(\*r), \hat\psi_{\beta\sigma'}\dg(\*r')\} = \delta_{\alpha\beta}\delta_{\sigma\sigma'}\delta(\*r-\*r')$. Finally, $n_g$ is an offset charge which is in principle gate controllable, but also unavoidably undergoes random fluctuations in any realistic setting.

The last two terms in~\cref{eq:Halpha} are BCS Hamiltonians for each quantum wire
\begin{equation}
\begin{aligned}
    \hat H_{e,\alpha} ={}& \sum_{\sigma=\uparrow,\downarrow} \int \dd^3 r \hat \psi_{\alpha\sigma}\dg(\*r) h_\alpha(\*r) \hat \psi_{\alpha\sigma}(\*r)\\
    + \int& \dd^3 r\left[\Delta \e^{-i\hat\varphi} \hat \psi_{\alpha\uparrow}\dg(\*r) \hat \psi_{\alpha\downarrow}\dg(\*r) + \hc \right].
\end{aligned}
\end{equation}
Here, the first line describes single-electron physics and the second line are superconducting pairing terms with a superconducting gap $\Delta$ and phase $\hat\varphi$, whose physical origin is proximity induced superconductivity due to contact with a nearby bulk superconductor. A physical realization of a quantum wire can be, \emph{e.g.}, a semiconducting nanowire with an epitaxially grown superconducting shell coating the wire.
The superconducting phase and electron number operator are canonical conjugate variables satisfying $[\hat N,\e^{i\hat \varphi/2}] = \e^{i\hat \varphi/2}$.
Crucially, we assume that both wires are in contact with a common conventional s-wave superconductor (illustrated with blue color in~\cref{fig:setup}) shunting the two wires such that the entire system behaves as a single superconducting island with a uniform superconducting phase. This configuration, first proposed in~\cite{Plugge2017}, has been dubbed a Majorana box qubit.

The single-electron Hamiltonian $h_\alpha(\*r)$ is of the generic form
\begin{equation}\label{eq:h0}
    h_\alpha(\*r) = \frac{-\hbar^2}{2m} \nabla^2 - eV(\*r) + \dots,
\end{equation}
where $m$ is the effective electron mass,
$V(\*r)$ refers to the total potential experienced by the electrons due to the confining potential of the nano-circuit as well as any (classical) gate voltages~\cite{Cottet15}. The trailing ellipses in~\cref{eq:h0} refer to spin-orbit coupling, Zeeman fields and any other single-electron physics necessary for the existence bound near-zero energy Majorana modes in the wire~\cite{Oreg2010,Lutchyn2010}. 
Here, we simply assume that each wire is in the topological regime with a pair of bound Majorana modes localized at the respective wire ends.

Coupling to the quantized electromagnetic field is introduced through an interaction Hamiltonian
\begin{equation}
    \hat H_{\text{tot}} = \hat H_\text{island} + \hat H_{r} + \hat H_{\text{island,int}},
\end{equation}
where $\hat H_r$ is the electromagnetic Hamiltonian, which in a single-mode approximation becomes $\hat H_r = \hbar\omega_r\hat a\dg \hat a$, with $\omega_r$ the resonator frequency and $\hat a$ the annihilation operator for the resonator mode.
The light-matter interaction describes capacitive coupling between the electrons on the island and the electric field of the resonator~\cite{Cottet15}
\begin{equation}\label{eq:Hint}
\begin{aligned}
\hat H_{\text{island,int}} ={}& 
i\hbar \lambda_C \hat N(\hat a\dg - \hat a)\\
-e \sum_{\alpha=L,R}& \sum_{\sigma=\uparrow,\downarrow} \int \dd^3 r \hat V_r(\*r) \hat \psi_{\alpha\sigma}\dg(\*r) \hat \psi_{\alpha\sigma}(\*r),
\end{aligned}
\end{equation}
where the coupling strength can be expressed as
\begin{equation}
    \lambda_C = -\omega_r\sqrt{\frac{\pi Z_r}{R_K}} \frac{C_{c}}{C_\text{island}},
\end{equation}
with $C_{c}$ the coupling capacitance between the island and the resonator, $C_\text{island} = e^2/(2E_{C})$ the island's capacitance, $Z_r$ the resonator's characteristic impedance and $R_K = h/e^2$ the quantum of resistance.
\Cref{eq:Hint} can be read as a quantized contribution to the gate voltage biasing the island, where in a single mode approximation the resonator voltage is
\begin{equation}
    \hat V_r(\*r) = i\omega_r \sqrt{\frac{\hbar Z_r}{2}} \frac{C_c}{C_\text{island}} u(\*r)(\hat a\dg - \hat a),
\end{equation}
with $u(\*r)$ a dimensionless resonator mode function describing the spatial dependence of the voltage biasing the island. 
Note that if the island is small compared to any spatial variation of the resonator mode function, we can take $u(\*r) \simeq 1$ and the Hamiltonian takes the simpler form $\hat H_\text{island,int} = i\hbar \lambda_C (\hat N + \hat n_e)(\hat a\dg - \hat a)$.

We are interested in the low-energy physics of $\hat H_{\text{island}}$. It is convenient to perform a unitary transformation~\cite{Hell16}
\begin{equation}\label{eq:U}
    \hat U = \e^{-i \hat\varphi \hat n_e/2},
\end{equation}
such that
\begin{align}
\hat U\dg \hat \psi_{\alpha\sigma}(\*r) \hat U ={}& \e^{-i\hat\varphi/2}\hat \psi_{\alpha\sigma}(\*r),\\
\hat U\dg \hat N \hat U ={}& \hat N - \hat n_e.
\end{align}
This unitary removes the dependence on $\hat \varphi$ from $\hat H_{e,\alpha}' = \hat U\dg \hat H_{e,\alpha} \hat U$ and $\hat n_e$ from the charging energy term in~\cref{eq:Halpha}.
The transformed $\hat H_{e,\alpha}'$ can then be diagonalized by an expansion of the electronic field in terms of Bogoliubov modes
\begin{equation}
\begin{aligned}
\hat\psi_{\alpha\sigma}(\*r) 
={}& \sum_k \left[ u_{\alpha \sigma k}(\*r) \hat b_{\alpha k} + v_{\alpha \sigma k}(\*r) \hat b_{\alpha k}\dg \right] \\
={}& f_{\alpha\sigma}(\*r)\gamma_{\alpha 1} + i g_{\alpha\sigma}(\*r)\gamma_{\alpha 2} + \dots,
\end{aligned}
\end{equation}
where $\hat b_{\alpha k}$ are Bogoliubov operators with associated Bogoliubov mode functions $\{u_{\alpha \sigma k}(\*r), v_{\alpha \sigma k}(\*r)\}$. In the second line above we have written the lowest energy Bogoliubov mode in terms of Majorana operators, $\hat b_{\alpha 0} = \half (\hat \gamma_{\alpha 1} + i\hat\gamma_{\alpha 2})$, and the ellipses refer to higher energy modes, $k=1,2,\dots$, which we drop in a low-energy approximation for each wire. Note that the Majorana mode functions $\{f_{\alpha\sigma}(\*r),g_{\alpha\sigma}(\*r)\}$ are real.

The resulting low-energy approximation to $\hat H_\text{island}' = \hat U\dg \hat H_\text{island} \hat U$ is
\begin{equation}
    \begin{aligned}
    \hat H_\text{island}' \simeq{}&
    E_{C}(\hat N - n_g)^2 
    + \sum_{\alpha=L,R}E_{\alpha} i\hat \gamma_{\alpha 1}\hat \gamma_{\alpha 2}.
    \end{aligned}
\end{equation}
Here
$E_\alpha$
is the energy splitting of the Majorana fermion, proportional to the Majorana mode function overlap which we assume to be exponentially small in the wire length $L_\alpha$, that is $E_{\alpha} \propto \e^{-L_\alpha/\xi_\alpha}$ with $\xi_\alpha$ a characteristic coherence length for the wire~\cite{Knapp2018}. We assume that $L_\alpha$ is large enough that we can set $E_\alpha=0$ from here on.

Similarly, the interaction Hamiltonian
$\hat H_\text{island,int}' = \hat U\dg \hat H_\text{island,int} \hat U$
is in the transformed frame given by
\begin{equation}
\begin{aligned}
\hat H_{\text{island,int}}' ={}& 
i\hbar \lambda_C \hat N(\hat a\dg - \hat a)\\
- \sum_{\alpha=L,R}& \sum_{\sigma=\uparrow,\downarrow} \int \dd^3 r \delta \mu(\*r) \hat \psi_{\alpha\sigma}\dg(\*r) \hat \psi_{\alpha\sigma}(\*r),
\end{aligned}
\end{equation}
where we have defined
\begin{equation}
\delta \mu(\*r) = \omega_r \sqrt{\frac{\pi Z_r}{R_K}} \frac{C_c}{C_\text{island}}[u_r(\*r)-1].
\end{equation}
We see that in the lumped element approximation mentioned above where we take $u(\*r) = 1$ across the entire island, the resonator voltage decouples from the unpaired electrons in this frame. However, it is worth noting that we can relax this assumption when the only relevant fermionic modes are the low-energy bound Majorana modes, since in the same low-energy approximation as before we have
\begin{equation}\label{eq:Hint2}
\begin{aligned}
&\hat H_{\text{island,int}}' \simeq{}
i\hbar \lambda_C \hat N(\hat a\dg - \hat a) \\
&+ \sum_{\alpha=L,R} i\lambda_\alpha i\hat\gamma_{\alpha 1}\hat\gamma_{\alpha 2}(\hat a\dg - \hat a)
+ i\hbar A(\hat a\dg - \hat a),
\end{aligned}
\end{equation}
where
\begin{align}
    &\begin{aligned}
    \lambda_{\alpha} = -2\omega_r \sqrt{\frac{\pi Z_r}{R_K}} \frac{C_c}{C_\text{island}}\sum_\sigma& \int \dd^3 r \left[u_r(\*r)-1\right] \\
    &\times f_{\alpha\sigma}(\*r) g_{\alpha\sigma}(\*r)\label{eq:lambda_alpha},
    \end{aligned}\\
    &\begin{aligned}
    A ={}& -\omega_r \sqrt{\frac{\pi Z_r}{R_K}} \frac{C_c}{C_\text{island}} \sum_{\alpha} \sum_\sigma
    \int \dd^3 r \left[u_r(\*r)-1\right] \\
    &\times [f_{\alpha\sigma}(\*r) f_{\alpha\sigma}(\*r) + g_{\alpha\sigma}(\*r) g_{\alpha\sigma}(\*r)].
    \end{aligned}
\end{align}
Crucially,  $\lambda_\alpha$ vanishes in the topological regime where the Majorana mode function overlap is exponentially small.  In the long wire limit we therefore take $\lambda_\alpha \to 0$.
Moreover, this is holds independently of the detailed form of $u(\*r)$, and we therefore expect the electromagnetic field to decouple from the Majorana modes even in a more general situation where the resonator mode function varies significantly over the qubit island. The term proportional to $A$ gives a small displacement of the resonator $\sim A/\omega_r$ which can be absorbed into a re-definition of $\hat a$. We simply ignore this term in the following.

The disappearance of $\hat n_e$ from the charging energy and the capacitive coupling to the resonator means that in the frame defined by~\cref{eq:U}, $\hat N$ effectively counts the \emph{total} charge on the island. This will become more clear in the next section when we introduce tunneling of electrons on and off the island.

\subsection{\label{sec:wiredotwire}TS-Sm-TS setup}

To be able to couple to pairs of Majorana operators such as $i\gamma_{L2}\gamma_{R1}$, where the two corresponding Majorana mode functions are localized on different wires, we introduce a tunable semiconducting region as a mediator, as illustrated in~\cref{fig:setup}. The Sm-segment is described by a set of electronic orbitals
\begin{equation}\label{eq:dot}
    \hat H_\text{B} = \sum_j h_j \hat b_j\dg \hat b_j + U\left( \sum_j \hat b_j\dg \hat b_j - n_b\right)^2
\end{equation}
where $\{\hat b_i,\hat b_j\dg\}=\delta_{ij}$,
and coupling between the wires and the semiconductor by a phenomenological tunneling Hamiltonian
\begin{equation}\label{eq:HT}
    \hat H_{T} = - \sum_{\alpha} \sum_j \sum_{\sigma=\uparrow,\downarrow} \int \dd^3 r t_{\alpha j\sigma}(\*r) \hat\psi_{\alpha\sigma}\dg(\*r) \hat b_j + \hc
\end{equation}
The Sm-segment also couples capacitively to the resonator field, described by a Hamiltonian~\cite{Cottet15}
\begin{equation}
    \hat H_{B,\text{int}} = \sum_j i\hbar \lambda_j \hat b_j\dg \hat b_j(\hat a\dg - \hat a),
\end{equation}
with $\lambda_j$ a coupling strength for Sm-orbital $j$.

Upon performing the unitary transformation~\cref{eq:U}
and the low-energy approximation as before,
one can write a total Hamiltonian for the system in the transformed frame
\begin{equation}\label{eq:Htot}
    \begin{aligned}
    \hat H_\text{tot}' ={}&
    \hat H_\text{island}'  + \hat H_{\text{island,int}}'
    + \hat H_B + \hat H_r \\
    &+ \hat H_{B,\text{int}}
    + \hat H_T'
    \end{aligned}
\end{equation}
where the low-energy approximation to the tunneling Hamiltonian is
\begin{equation}\label{eq:HT2}
    \begin{aligned}
    \hat H_T' \simeq \sum_j \Big[&\frac{i t_{Lj}}{2} \e^{i\hat\varphi/2}\hat \gamma_{L2}\hat b_j \\
    & - \frac{t_{Rj}}{2} \e^{i\hat\varphi/2} \hat \gamma_{R1} \hat b_j
    + \hc \Big],
    \end{aligned}
\end{equation}
with
\begin{subequations}\label{eq:tunnelcouplings}
\begin{align}
    t_{Lj} ={}& 2\sum_{\sigma=\uparrow,\downarrow} \int \dd^3 r t_{Lj\sigma}(\*r)g_{L\sigma}(\*r),\\
    t_{Rj} ={}& 2\sum_{\sigma=\uparrow,\downarrow} \int \dd^3 r t_{Rj\sigma}(\*r) f_{R\sigma}(\*r).
\end{align}
\end{subequations}
We have here assumed that the overlap between the tunnel couplings $t_{\alpha j}(\*r)$ and the Majorana mode functions at the far ends of the wires, \emph{i.e.}, $f_{L\sigma}(\*r)$ and $g_{R\sigma}(\*r)$, is negligible.

The form of $\hat H_T'$ clarifies the role of $\hat N$ in the frame defined by~\cref{eq:U}. The operator $\e^{i\hat\varphi/2}$ increases $\hat N$ by one, \emph{i.e}, $\e^{i\hat\varphi/2}\ket{N} = \ket{N+1}$ for $\ket{N}$ an eigenstate of $\hat N$ with eigenvalue $N$. By introducing 
Majorana fermion operators for each wire $\hat f_L = \half(\hat \gamma_{L1} + i \hat \gamma_{L2})$, $\hat f_R = \half(\hat \gamma_{R1} + i \hat \gamma_{R2})$, such that
\begin{align}
\hat\gamma_{L2} ={}& i\hat f_L\dg - i \hat f_L,\\
\hat\gamma_{R1} ={}& \hat f_R\dg + \hat f_R,
\end{align}
we see that the action of an operator like $\e^{i\hat\varphi/2} \hat\gamma_{L2} \hat b_j$ is to remove one electron from the semiconductor, increase $\hat N$ by one, and flip the state of the fermion associated to $\hat f_L$ on the left wire. Thus, in this frame, $\hat N$ counts the total charge.

\subsection{\label{app:diagonalization}Diagonalizing the TS-Sm-TS Hamiltonian}
In the proposed readout protocol, the coupling to the semiconducting barrier is gradually turned on such that the initial near-zero energy logical qubit eigenstates evolve into hybridized states partially localized in the semiconducting segment. The logical states then become split in energy and couple to the resonator field. The key physics can be exposed by diagonalizing the Hamiltonian for the TS-Sm-TS system, excluding the coupling to the resonator, \emph{i.e.}, 
\begin{equation}\label{eq:H_wdw}
\hat H_\text{q}' = \hat H_\text{island}' + \hat H_B' + \hat H_T'.
\end{equation}
We only consider a single Sm-orbital $j=0$ in this section, for simplicity.

It is convenient to first combine $\hat \gamma_{L2}$ and $\hat \gamma_{R1}$ in a single fermion $\hat f = \half(\hat \gamma_{L2} + i\hat\gamma_{R1})$ such that
the tunneling Hamiltonian can be written
\begin{equation}\label{eq:HT}
    \hat H_T' = \frac{i t_- }{2} \e^{i\hat\varphi/2}\hat f\dg  \hat b_0
     + \frac{i t_+}{2} \e^{i\hat\varphi/2} \hat f\hat b_0
    + \hc,
\end{equation}
where we have defined $t_\pm = t_L\e^{i\varphi_x/2} \pm t_R$.
We here take $t_{L}$ and $t_{R}$ to be real and positive without loss of generality. Flux quantization around the loop formed by the superconducting island and TS-Sm-TS junction (see~\cref{fig:setup}) is accounted for by including the external flux contribution $\varphi_x = \Phi_x/\Phi_0$, with $\Phi_0 = h/2e$ the flux quantum, in the tunneling amplitudes $t_\pm$.

To diagonalize~\cref{eq:H_wdw}, we first note that the Hamiltonian only induces transitions within a two-level subspace \{$\ket{0_n} \equiv \ket{N=n, n_0=0}, \ket{1_n} \equiv \ket{N=n-1, n_0=1}\}$ of the island-charge/dot subsystem, where $n_0$ denotes the occupancy of the dot. We can thus treat each subspace labeled by $n$ independently. 

First define a new lowering operator
\begin{equation}
    \hat c_n = \ket{0_n}\bra{1_n},
\end{equation}
such that in the $\{\ket{0_n},\ket{1_n}\}$ subspace the Hamiltonian becomes
\begin{equation}\label{eq:Hq_block}
    \begin{aligned}
    \hat H_{q,n}' ={}& \delta(n) \hat c_n\dg \hat c_n  + E_C(n-n_g)^2\\
    &+ \left(\frac{i t_- }{2} \hat f\dg  \hat c_n
    + \frac{i t_+}{2} \hat f\hat c_n + \hc\right),
    \end{aligned}
\end{equation}
where $\delta(n) = \varepsilon_0 - 2E_C(n-n_g) + E_C$.
This quadratic Hamiltonian can be diagonalized exactly by a unitary transformation $\hat U_n = \e^{-\hat S_n}$ with
\begin{equation}
    \hat S_n = \alpha_-^{(n)} \hat c_n\dg \hat f  - \alpha_+^{(n)}\hat c_n\dg f\dg  - \hc
\end{equation}
With the choice
\begin{equation}
    \tan(2|\alpha_\pm^{(n)}|) = \frac{|t_\pm|}{\delta(n)},
    \quad \frac{\alpha_\pm^{(n)}}{|\alpha_\pm^{(n)}|} = \frac{-it_\pm^*}{|t_\pm|},
\end{equation}
we find
\begin{equation}
    \begin{aligned}
    \hat H''_{q,n} ={}& \hat U_n\dg \hat H_{q,n}' \hat U_n\\
    ={}& \varepsilon_{c}(n) \hat c_n\dg \hat c_n + \varepsilon_{f}(n)\hat f\dg \hat f + E(n)
    \end{aligned}
\end{equation}
with
\begin{align}
\varepsilon_{c}(n) ={}& \frac{\sgn\delta(n)}{2}\left[f_+(n) + f_-(n)\right],\\
\varepsilon_{f}(n) ={}& \frac{\sgn\delta(n)}{2}\left[f_+(n) - f_-(n)\right],\\
E(n) ={}& E_C(n-n_g)^2 + \half[\delta(n)-\varepsilon_c(n)-\varepsilon_f(n)],
\end{align}
For notational convenience we have defined $f_\pm(n) = \sqrt{\delta(n)^2 +  t_L^2 + t_R^2 \pm 2t_Lt_R\cos\left(\frac{\varphi_x}{2}\right)}$.

To consider the coupling to the resonator we need to also transform the interaction Hamiltonian.
The total interaction Hamiltonian in the $n$th subspace is
\begin{equation}
    \hat H_{\text{int},n}' = i\hbar(\lambda_0-\lambda_C)\hat c_n\dg \hat c_n(\hat a\dg - \hat a)
    + i\hbar n \lambda_C (\hat a\dg - \hat a),\\
\end{equation}
which transforms to
\begin{equation}\label{eq:app:Hintexact}
\begin{aligned}
\hat H&_{\text{int},n}'' = i\hbar \left[g_{c}(n)\hat c_n\dg c_n + g_{f}(n) \hat f\dg f\right](\hat a\dg - \hat a) \\
+& i \hbar\left[w_{-}(n) \hat f\dg \hat c_n + w_{+}(n) \hat f \hat c_n + \hc \right](\hat a\dg - \hat a)\\
+& i\hbar n \lambda_C (\hat a\dg - \hat a),
\end{aligned}
\end{equation}
with
\begin{align}
g_{c}(n) ={}& \frac{\lambda_0-\lambda_C}{2}\left(\frac{\delta(n)}{f_-(n)} + \frac{\delta(n)}{f_+(n)} \right),\\
g_{f}(n) ={}& \frac{\lambda_C-\lambda_0}{2}\left(\frac{\delta(n)}{f_-(n)} - \frac{\delta(n)}{f_+(n)} \right),\\
w_{\pm}(n) ={}& (\lambda_0-\lambda_C)\frac{it_\pm}{2\delta(n)}.
\end{align}

The advantage of this change of frame is that all terms that are off-diagonal in the electron operators are now of order $|\hbar w_\pm(n)|$. 
We can treat $\hbar w_\pm(n)$ as small parameters, assuming that the relevant electronic transition are all far detuned from the resonator energy.
We neglect these terms from now on, which is valid as long as $\delta_\lambda=\lambda_0-\lambda_C$ is sufficiently small compared to the energy cost $\delta(n)$ of moving an electron from the island to the barrier orbital.

Resumming, the Hamiltonians $\hat H_q'' = \sum_n \hat H_{q,n}''\hat P_n$ and $\hat H_\text{int}'' = \sum_n \hat H_{\text{int},n}'' \hat P_n$, where $\hat P_n$ is a projector onto the $\{\ket{0_n},\ket{1_n}\}$ subspace, can be written
\begin{equation}
    \begin{aligned}
    \hat H''_q &= \varepsilon_{c}(\hat N+1) \hat b_0\dg \hat b_0
    + \varepsilon_f(\hat N)\hat f\dg \hat f \\
    &+ [\varepsilon_f(\hat N+1)-\varepsilon_f(\hat N))]\hat b_0\dg \hat b_0\hat f\dg \hat f \\
    &+ E(\hat N) + [E(\hat N+1)- E(\hat N)]\hat b_0\dg \hat b_0,
    \end{aligned}
\end{equation}
and
\begin{equation}
    \begin{aligned}
    \hat H_\text{int}'' &\simeq
    i\hbar [g_c(\hat N+1) + \lambda_C] \hat b_0\dg \hat b_0(\hat a\dg-\hat a) \\
    &+ i\hbar g_f(\hat N)\hat f\dg \hat f (\hat a\dg - \hat a)\\
    &+ i\hbar [g_f(\hat N+1)-g_f(\hat N))]\hat b_0\dg \hat b_0\hat f\dg \hat f(\hat a\dg - \hat a) \\
    &+ i\hbar \lambda_C \hat N (\hat a\dg - \hat a),
    \end{aligned}
\end{equation}
where the various functions of $\hat N$ are diagonal operators in the charge basis defined through $f(\hat N+a) = \sum_{n} f(n+a)\ket n\bra n$. We emphasize that the diagonalization of $\hat H_q''$ is exact, and the only approximation is made in $\hat H_\text{int}''$.

Since $\hat H_\text{tot}'' = \hat H_q'' + \hat H_\text{int}'' + \hat H_r$ conserves the charge number and Sm-occupation at this level of approximation, we can assume that the charge and Sm-degrees of freedom remain in a definite state, and replace $\hat N\to N$, $\hat b_0\dg b_0 \to n_0$.
This amounts to an ``adiabatic elimination'' of the Sm and island charge degrees of freedom.
In particular, for $n_g \simeq 0$ and large $E_C$, $\varepsilon_0$, we can assume the charge subsystem to be in the state $\ket{N=0,n_0=0}$ and drop terms proportional to $\hat N$ and $\hat b_0\dg b_0$ in $\hat H''$.
The Hamiltonian in~\cref{eq:Hpert} is then finally found by defining $\hat \sigma_z = 2\hat f\dg \hat f - 1$.
\begin{equation}\label{eq:Hpert2}
    \hat H_\text{tot}'' \simeq
    \hbar \omega_r \hat a\dg \hat a
    + \frac{\hbar \omega_q}{2} \hat \sigma_z
    + i\hbar g_z (\hat \sigma_z+1) (\hat a\dg - \hat a)
\end{equation}
with $\hbar\omega_q = \varepsilon_{f}(0)$, $g_z = g_{f}(0)/2$, and we have dropped a constant term.

It is also insightful to consider approximate expressions for the parameters in~\cref{eq:Hpert} in the limit $t_L, t_R \ll \delta$, where $\delta\equiv\delta(0)=E_C+\varepsilon_0$. The expressions greatly simplify in this limit, which can be useful to gain physical insight, but we emphasize that the readout protocol is not limited to this regime. Indeed, a strong hybridization of the Sm-orbitals and Majorana modes might be preferable. The relevant parameters are in the small tunneling regime approximated by
\begin{align}
    \hbar\omega_q \simeq{}& \frac{t_Lt_R\cos\left(\varphi_x/2\right)}{\delta(n)},\\
    \hbar g_z \simeq{}& \frac{\lambda_C-\lambda_0}{4}\frac{t_Lt_R\cos\left(\varphi_x/2\right)}{\delta(n)^2}.
\end{align}

Note that although~\cref{eq:Hpert2} is perturbative in the coupling $\delta_\lambda = \lambda_0-\lambda_C$, higher order terms in a Schrieffer-Wolff~\cite{bravyi2011schrieffer} expansion of~\cref{eq:app:Hintexact} will be proportional to $\hat f\dg \hat f$ or the identity, and thus still commute with $\hat \sigma_z$. 
Since it is advantageous to have fairly large $\{t_L,t_R\}/\delta$  this suggests a regime where both the tunneling rates and $\delta$ are large compared to $\delta_\lambda$. For $\delta_\lambda/(2\pi)$ in the $10$--$100$ MHz range, $\delta/\hbar$ and $\{t_L,t_R\}/\hbar$ in the 10 GHz range or more is desirable~\cite{fasth2005tunable,Mukhopadhyay20182}. Another constraint is related to the charging energy of the island, $E_C$. Protection from quasi-particle poisoning favors a large $E_C$, but this constraints $\delta_\lambda$ since the island capacitance can not be too large. The coupling strength to the resonator can, however, be boosted using a high-impedance resonator, as has recently been used to achieve strong spin-photon coupling~\cite{Stockklauser2017,Mi2018,Samkharadze2018}.

\section{\label{app:env}Coupling to the environment}

Noise in system parameters such as $n_g$, $t_{\alpha j}$, $\varepsilon_j$ or $\varphi_x$, leads to fluctuations in $\omega_q$ and $g_z$, but preserves the general form of~\cref{eq:Hpert}. This type of noise might reduce the signal-to-noise ratio in a measurement, but does not cause state-transitions, \emph{i.e.}, it does not change the QND nature of the readout.

Understanding noise due to coupling to the surrounding electromagnetic environment is also straight forward in the sense that there is nothing special about the resonator mode that was singled out in~\cref{app:interaction}. In other words, other modes of the electromagnetic field couples in exactly the same way, such that in the same low-energy approximation as before, we expect the coupling to the environment to be of the form
\begin{equation}\label{eq:H_env_int}
    H_\text{env,int} = \lambda_{ij}^B \hat \gamma_i \hat \gamma_j \hat B(t),
\end{equation}
for any pair of Majorana modes $\hat \gamma_i,\hat \gamma_j$.
Here $\hat B(t)$ could include both classical stochastic processes, describing noise in system parameters, and quantum noise through a bath operator of the generic form~\cite{Gardiner2004}
\begin{equation}\label{eq:Bt}
    \hat B(t) = i\int_0^\infty \dd\omega \lambda_\omega (\hat b_\omega\dg \e^{i\omega t} - \hc),
\end{equation}
where $\lambda_\omega$ are coupling constants and $[\hat b_\omega,\hat b_{\omega'}\dg] = \delta(\omega-\omega')$.
Crucially the coupling $\lambda_{ij}^B$ in~\cref{eq:H_env_int} is proportional to
\begin{equation}\label{eq:lambda_ijB}
    \lambda_{ij}^B \propto \int \dd^3 r f_i(\*r) g_j(\*r),
\end{equation}
where $f_i(\*r)$ and $g_j(\*r)$ are the corresponding Majorana wavefunctions. For two Majorana modes localized on the same superconducting island, this follows from the derivation leading up to~\cref{eq:Hint2,eq:lambda_alpha} by replacing the single-mode resonator field by the multi-mode field~\cref{eq:Bt}. For two Majorana modes tunnel coupled to a common non-superconducting region as in~\cref{sec:wiredotwire}, the form~\cref{eq:lambda_ijB} follows since the tunnel coupling are proportional to the overlap of the corresponding Majorana wavefunctions and the electronic orbitals in the non-superconducting region, c.f.~\cref{eq:HT2,eq:tunnelcouplings}.

In summary, any coupling to the surrounding electromagnetic environment is exponentially suppressed for Majorana modes that are far apart, due to the fractional and non-local nature of the Majorana modes. We can therefore choose to couple only to a single operator $i\hat \gamma_i \hat \gamma_j$, with coupling to other Majorana modes highly suppressed. This observable is thus a constant of motion (up to exponential corrections in the wire lengths and the usual arguments about a sufficiently large superconducting gap, etc.). In this sense the QND nature of the measurement is topologically protected.

\section{\label{app:fidelity}Measurement fidelity for longitudinal readout}

The simple form of~\cref{eq:Hpert} of the main paper makes it possible to find an analytical expression for the measurement fidelity, which is defined as
\begin{equation}
    F = \half\left[p(1|1) + p(-1|-1)\right] = p(1|1),
\end{equation}
where $p(i|j)$ is the probability of assigning outcome $i$ given $\hat \sigma_z = j$ (note that $\hat \sigma_z$ is a constant of motion), and by symmetry $p(1|1) = p(-1|-1)$ in the current context. Using the results from the Supplementary Material of Ref.~\cite{Didier2015} we have that a homodyne measurement of the resonator's output field is described by a Gaussian random variable with mean
\begin{equation}
    \mu_{\pm 1} = \pm \mu = \pm 2|\tilde g_z|\tau \left[1 - \frac{2}{\kappa t}\left(1-\e^{-\kappa\tau/2}\right)\right],
\end{equation}
and variance $\sigma_{\pm 1} = \kappa \tau$, where $\tilde g_z$ is the amplitude of the modulation of the longitudinal coupling and $\tau$ is the measurement time. The $\pm 1$ notation refers to the two values $\hat \sigma_z = \pm 1$.
The probability $F = p(1|1)$ is then easily found
\begin{equation}
    \begin{aligned}
    F ={}& \frac{1}{\sqrt{2\pi}\sigma} \int_0^\infty \dd x \e^{-(x-\mu)^2/2\sigma^2}\\
    ={}& 1 - \half \text{erfc}(\frac{\mu}{\sqrt 2 \sigma}).
    \end{aligned}
\end{equation}


\begin{thebibliography}{38}%
\makeatletter
\providecommand \@ifxundefined [1]{%
 \@ifx{#1\undefined}
}%
\providecommand \@ifnum [1]{%
 \ifnum #1\expandafter \@firstoftwo
 \else \expandafter \@secondoftwo
 \fi
}%
\providecommand \@ifx [1]{%
 \ifx #1\expandafter \@firstoftwo
 \else \expandafter \@secondoftwo
 \fi
}%
\providecommand \natexlab [1]{#1}%
\providecommand \enquote  [1]{``#1''}%
\providecommand \bibnamefont  [1]{#1}%
\providecommand \bibfnamefont [1]{#1}%
\providecommand \citenamefont [1]{#1}%
\providecommand \href@noop [0]{\@secondoftwo}%
\providecommand \href [0]{\begingroup \@sanitize@url \@href}%
\providecommand \@href[1]{\@@startlink{#1}\@@href}%
\providecommand \@@href[1]{\endgroup#1\@@endlink}%
\providecommand \@sanitize@url [0]{\catcode `\\12\catcode `\$12\catcode
  `\&12\catcode `\#12\catcode `\^12\catcode `\_12\catcode `\%12\relax}%
\providecommand \@@startlink[1]{}%
\providecommand \@@endlink[0]{}%
\providecommand \url  [0]{\begingroup\@sanitize@url \@url }%
\providecommand \@url [1]{\endgroup\@href {#1}{\urlprefix }}%
\providecommand \urlprefix  [0]{URL }%
\providecommand \Eprint [0]{\href }%
\providecommand \doibase [0]{http://dx.doi.org/}%
\providecommand \selectlanguage [0]{\@gobble}%
\providecommand \bibinfo  [0]{\@secondoftwo}%
\providecommand \bibfield  [0]{\@secondoftwo}%
\providecommand \translation [1]{[#1]}%
\providecommand \BibitemOpen [0]{}%
\providecommand \bibitemStop [0]{}%
\providecommand \bibitemNoStop [0]{.\EOS\space}%
\providecommand \EOS [0]{\spacefactor3000\relax}%
\providecommand \BibitemShut  [1]{\csname bibitem#1\endcsname}%
\let\auto@bib@innerbib\@empty
\bibitem [{\citenamefont {Mourik}\ \emph {et~al.}(2012)\citenamefont {Mourik},
  \citenamefont {Zuo}, \citenamefont {Frolov}, \citenamefont {Plissard},
  \citenamefont {Bakkers},\ and\ \citenamefont {Kouwenhoven}}]{Mourik2012}%
  \BibitemOpen
  \bibfield  {author} {\bibinfo {author} {\bibfnamefont {V.}~\bibnamefont
  {Mourik}}, \bibinfo {author} {\bibfnamefont {K.}~\bibnamefont {Zuo}},
  \bibinfo {author} {\bibfnamefont {S.~M.}\ \bibnamefont {Frolov}}, \bibinfo
  {author} {\bibfnamefont {S.}~\bibnamefont {Plissard}}, \bibinfo {author}
  {\bibfnamefont {E.~P.}\ \bibnamefont {Bakkers}}, \ and\ \bibinfo {author}
  {\bibfnamefont {L.~P.}\ \bibnamefont {Kouwenhoven}},\ }\href@noop {}
  {\bibfield  {journal} {\bibinfo  {journal} {Science}\ }\textbf {\bibinfo
  {volume} {336}},\ \bibinfo {pages} {1003} (\bibinfo {year}
  {2012})}\BibitemShut {NoStop}%
\bibitem [{\citenamefont {Deng}\ \emph {et~al.}(2016)\citenamefont {Deng},
  \citenamefont {Vaitiek{\.e}nas}, \citenamefont {Hansen}, \citenamefont
  {Danon}, \citenamefont {Leijnse}, \citenamefont {Flensberg}, \citenamefont
  {Nyg{\aa}rd}, \citenamefont {Krogstrup},\ and\ \citenamefont
  {Marcus}}]{Deng2016}%
  \BibitemOpen
  \bibfield  {author} {\bibinfo {author} {\bibfnamefont {M.}~\bibnamefont
  {Deng}}, \bibinfo {author} {\bibfnamefont {S.}~\bibnamefont
  {Vaitiek{\.e}nas}}, \bibinfo {author} {\bibfnamefont {E.~B.}\ \bibnamefont
  {Hansen}}, \bibinfo {author} {\bibfnamefont {J.}~\bibnamefont {Danon}},
  \bibinfo {author} {\bibfnamefont {M.}~\bibnamefont {Leijnse}}, \bibinfo
  {author} {\bibfnamefont {K.}~\bibnamefont {Flensberg}}, \bibinfo {author}
  {\bibfnamefont {J.}~\bibnamefont {Nyg{\aa}rd}}, \bibinfo {author}
  {\bibfnamefont {P.}~\bibnamefont {Krogstrup}}, \ and\ \bibinfo {author}
  {\bibfnamefont {C.~M.}\ \bibnamefont {Marcus}},\ }\href@noop {} {\bibfield
  {journal} {\bibinfo  {journal} {Science}\ }\textbf {\bibinfo {volume}
  {354}},\ \bibinfo {pages} {1557} (\bibinfo {year} {2016})}\BibitemShut
  {NoStop}%
\bibitem [{\citenamefont {Albrecht}\ \emph {et~al.}(2016)\citenamefont
  {Albrecht}, \citenamefont {Higginbotham}, \citenamefont {Madsen},
  \citenamefont {Kuemmeth}, \citenamefont {Jespersen}, \citenamefont
  {Nyg{\aa}rd}, \citenamefont {Krogstrup},\ and\ \citenamefont
  {Marcus}}]{Albrecht2016}%
  \BibitemOpen
  \bibfield  {author} {\bibinfo {author} {\bibfnamefont {S.~M.}\ \bibnamefont
  {Albrecht}}, \bibinfo {author} {\bibfnamefont {A.}~\bibnamefont
  {Higginbotham}}, \bibinfo {author} {\bibfnamefont {M.}~\bibnamefont
  {Madsen}}, \bibinfo {author} {\bibfnamefont {F.}~\bibnamefont {Kuemmeth}},
  \bibinfo {author} {\bibfnamefont {T.~S.}\ \bibnamefont {Jespersen}}, \bibinfo
  {author} {\bibfnamefont {J.}~\bibnamefont {Nyg{\aa}rd}}, \bibinfo {author}
  {\bibfnamefont {P.}~\bibnamefont {Krogstrup}}, \ and\ \bibinfo {author}
  {\bibfnamefont {C.}~\bibnamefont {Marcus}},\ }\href@noop {} {\bibfield
  {journal} {\bibinfo  {journal} {Nature}\ }\textbf {\bibinfo {volume} {531}},\
  \bibinfo {pages} {206} (\bibinfo {year} {2016})}\BibitemShut {NoStop}%
\bibitem [{\citenamefont {Nichele}\ \emph {et~al.}(2017)\citenamefont
  {Nichele}, \citenamefont {Drachmann}, \citenamefont {Whiticar}, \citenamefont
  {O'Farrell}, \citenamefont {Suominen}, \citenamefont {Fornieri},
  \citenamefont {Wang}, \citenamefont {Gardner}, \citenamefont {Thomas},
  \citenamefont {Hatke}, \citenamefont {Krogstrup}, \citenamefont {Manfra},
  \citenamefont {Flensberg},\ and\ \citenamefont {Marcus}}]{Nichele2017}%
  \BibitemOpen
  \bibfield  {author} {\bibinfo {author} {\bibfnamefont {F.}~\bibnamefont
  {Nichele}}, \bibinfo {author} {\bibfnamefont {A.~C.~C.}\ \bibnamefont
  {Drachmann}}, \bibinfo {author} {\bibfnamefont {A.~M.}\ \bibnamefont
  {Whiticar}}, \bibinfo {author} {\bibfnamefont {E.~C.~T.}\ \bibnamefont
  {O'Farrell}}, \bibinfo {author} {\bibfnamefont {H.~J.}\ \bibnamefont
  {Suominen}}, \bibinfo {author} {\bibfnamefont {A.}~\bibnamefont {Fornieri}},
  \bibinfo {author} {\bibfnamefont {T.}~\bibnamefont {Wang}}, \bibinfo {author}
  {\bibfnamefont {G.~C.}\ \bibnamefont {Gardner}}, \bibinfo {author}
  {\bibfnamefont {C.}~\bibnamefont {Thomas}}, \bibinfo {author} {\bibfnamefont
  {A.~T.}\ \bibnamefont {Hatke}}, \bibinfo {author} {\bibfnamefont
  {P.}~\bibnamefont {Krogstrup}}, \bibinfo {author} {\bibfnamefont {M.~J.}\
  \bibnamefont {Manfra}}, \bibinfo {author} {\bibfnamefont {K.}~\bibnamefont
  {Flensberg}}, \ and\ \bibinfo {author} {\bibfnamefont {C.~M.}\ \bibnamefont
  {Marcus}},\ }\href {\doibase 10.1103/PhysRevLett.119.136803} {\bibfield
  {journal} {\bibinfo  {journal} {Phys. Rev. Lett.}\ }\textbf {\bibinfo
  {volume} {119}},\ \bibinfo {pages} {136803} (\bibinfo {year}
  {2017})}\BibitemShut {NoStop}%
\bibitem [{\citenamefont {Zhang}\ \emph {et~al.}(2018)\citenamefont {Zhang},
  \citenamefont {Liu}, \citenamefont {Gazibegovic}, \citenamefont {Xu},
  \citenamefont {Logan}, \citenamefont {Wang}, \citenamefont {Van~Loo},
  \citenamefont {Bommer}, \citenamefont {De~Moor}, \citenamefont {Car} \emph
  {et~al.}}]{Zhang2018}%
  \BibitemOpen
  \bibfield  {author} {\bibinfo {author} {\bibfnamefont {H.}~\bibnamefont
  {Zhang}}, \bibinfo {author} {\bibfnamefont {C.-X.}\ \bibnamefont {Liu}},
  \bibinfo {author} {\bibfnamefont {S.}~\bibnamefont {Gazibegovic}}, \bibinfo
  {author} {\bibfnamefont {D.}~\bibnamefont {Xu}}, \bibinfo {author}
  {\bibfnamefont {J.~A.}\ \bibnamefont {Logan}}, \bibinfo {author}
  {\bibfnamefont {G.}~\bibnamefont {Wang}}, \bibinfo {author} {\bibfnamefont
  {N.}~\bibnamefont {Van~Loo}}, \bibinfo {author} {\bibfnamefont {J.~D.}\
  \bibnamefont {Bommer}}, \bibinfo {author} {\bibfnamefont {M.~W.}\
  \bibnamefont {De~Moor}}, \bibinfo {author} {\bibfnamefont {D.}~\bibnamefont
  {Car}},  \emph {et~al.},\ }\href@noop {} {\bibfield  {journal} {\bibinfo
  {journal} {Nature}\ }\textbf {\bibinfo {volume} {556}},\ \bibinfo {pages}
  {74} (\bibinfo {year} {2018})}\BibitemShut {NoStop}%
\bibitem [{\citenamefont {Suominen}\ \emph {et~al.}(2017)\citenamefont
  {Suominen}, \citenamefont {Kjaergaard}, \citenamefont {Hamilton},
  \citenamefont {Shabani}, \citenamefont {Palmstr\o{}m}, \citenamefont
  {Marcus},\ and\ \citenamefont {Nichele}}]{Suominen2017}%
  \BibitemOpen
  \bibfield  {author} {\bibinfo {author} {\bibfnamefont {H.~J.}\ \bibnamefont
  {Suominen}}, \bibinfo {author} {\bibfnamefont {M.}~\bibnamefont
  {Kjaergaard}}, \bibinfo {author} {\bibfnamefont {A.~R.}\ \bibnamefont
  {Hamilton}}, \bibinfo {author} {\bibfnamefont {J.}~\bibnamefont {Shabani}},
  \bibinfo {author} {\bibfnamefont {C.~J.}\ \bibnamefont {Palmstr\o{}m}},
  \bibinfo {author} {\bibfnamefont {C.~M.}\ \bibnamefont {Marcus}}, \ and\
  \bibinfo {author} {\bibfnamefont {F.}~\bibnamefont {Nichele}},\ }\href
  {\doibase 10.1103/PhysRevLett.119.176805} {\bibfield  {journal} {\bibinfo
  {journal} {Phys. Rev. Lett.}\ }\textbf {\bibinfo {volume} {119}},\ \bibinfo
  {pages} {176805} (\bibinfo {year} {2017})}\BibitemShut {NoStop}%
\bibitem [{\citenamefont {Bonderson}\ \emph {et~al.}(2008)\citenamefont
  {Bonderson}, \citenamefont {Freedman},\ and\ \citenamefont
  {Nayak}}]{Bonderson08}%
  \BibitemOpen
  \bibfield  {author} {\bibinfo {author} {\bibfnamefont {P.}~\bibnamefont
  {Bonderson}}, \bibinfo {author} {\bibfnamefont {M.}~\bibnamefont {Freedman}},
  \ and\ \bibinfo {author} {\bibfnamefont {C.}~\bibnamefont {Nayak}},\
  }\href@noop {} {\bibfield  {journal} {\bibinfo  {journal} {Phys. Rev. Lett.}\
  }\textbf {\bibinfo {volume} {101}},\ \bibinfo {pages} {010501} (\bibinfo
  {year} {2008})}\BibitemShut {NoStop}%
\bibitem [{\citenamefont {Plugge}\ \emph {et~al.}(2017)\citenamefont {Plugge},
  \citenamefont {Rasmussen}, \citenamefont {Egger},\ and\ \citenamefont
  {Flensberg}}]{Plugge2017}%
  \BibitemOpen
  \bibfield  {author} {\bibinfo {author} {\bibfnamefont {S.}~\bibnamefont
  {Plugge}}, \bibinfo {author} {\bibfnamefont {A.}~\bibnamefont {Rasmussen}},
  \bibinfo {author} {\bibfnamefont {R.}~\bibnamefont {Egger}}, \ and\ \bibinfo
  {author} {\bibfnamefont {K.}~\bibnamefont {Flensberg}},\ }\href@noop {}
  {\bibfield  {journal} {\bibinfo  {journal} {New J. Phys.}\ }\textbf {\bibinfo
  {volume} {19}},\ \bibinfo {pages} {012001} (\bibinfo {year}
  {2017})}\BibitemShut {NoStop}%
\bibitem [{\citenamefont {Karzig}\ \emph {et~al.}(2017)\citenamefont {Karzig},
  \citenamefont {Knapp}, \citenamefont {Lutchyn}, \citenamefont {Bonderson},
  \citenamefont {Hastings}, \citenamefont {Nayak}, \citenamefont {Alicea},
  \citenamefont {Flensberg}, \citenamefont {Plugge}, \citenamefont {Oreg} \emph
  {et~al.}}]{Karzig2017}%
  \BibitemOpen
  \bibfield  {author} {\bibinfo {author} {\bibfnamefont {T.}~\bibnamefont
  {Karzig}}, \bibinfo {author} {\bibfnamefont {C.}~\bibnamefont {Knapp}},
  \bibinfo {author} {\bibfnamefont {R.~M.}\ \bibnamefont {Lutchyn}}, \bibinfo
  {author} {\bibfnamefont {P.}~\bibnamefont {Bonderson}}, \bibinfo {author}
  {\bibfnamefont {M.~B.}\ \bibnamefont {Hastings}}, \bibinfo {author}
  {\bibfnamefont {C.}~\bibnamefont {Nayak}}, \bibinfo {author} {\bibfnamefont
  {J.}~\bibnamefont {Alicea}}, \bibinfo {author} {\bibfnamefont
  {K.}~\bibnamefont {Flensberg}}, \bibinfo {author} {\bibfnamefont
  {S.}~\bibnamefont {Plugge}}, \bibinfo {author} {\bibfnamefont
  {Y.}~\bibnamefont {Oreg}},  \emph {et~al.},\ }\href@noop {} {\bibfield
  {journal} {\bibinfo  {journal} {Phys. Rev. B}\ }\textbf {\bibinfo {volume}
  {95}},\ \bibinfo {pages} {235305} (\bibinfo {year} {2017})}\BibitemShut
  {NoStop}%
\bibitem [{\citenamefont {Barends}\ \emph {et~al.}(2014)\citenamefont
  {Barends}, \citenamefont {Kelly}, \citenamefont {Megrant}, \citenamefont
  {Veitia}, \citenamefont {Sank}, \citenamefont {Jeffrey}, \citenamefont
  {White}, \citenamefont {Mutus}, \citenamefont {Fowler}, \citenamefont
  {Campbell} \emph {et~al.}}]{Barends2014}%
  \BibitemOpen
  \bibfield  {author} {\bibinfo {author} {\bibfnamefont {R.}~\bibnamefont
  {Barends}}, \bibinfo {author} {\bibfnamefont {J.}~\bibnamefont {Kelly}},
  \bibinfo {author} {\bibfnamefont {A.}~\bibnamefont {Megrant}}, \bibinfo
  {author} {\bibfnamefont {A.}~\bibnamefont {Veitia}}, \bibinfo {author}
  {\bibfnamefont {D.}~\bibnamefont {Sank}}, \bibinfo {author} {\bibfnamefont
  {E.}~\bibnamefont {Jeffrey}}, \bibinfo {author} {\bibfnamefont {T.~C.}\
  \bibnamefont {White}}, \bibinfo {author} {\bibfnamefont {J.}~\bibnamefont
  {Mutus}}, \bibinfo {author} {\bibfnamefont {A.~G.}\ \bibnamefont {Fowler}},
  \bibinfo {author} {\bibfnamefont {B.}~\bibnamefont {Campbell}},  \emph
  {et~al.},\ }\href@noop {} {\bibfield  {journal} {\bibinfo  {journal}
  {Nature}\ }\textbf {\bibinfo {volume} {508}},\ \bibinfo {pages} {500}
  (\bibinfo {year} {2014})}\BibitemShut {NoStop}%
\bibitem [{\citenamefont {Walter}\ \emph {et~al.}(2017)\citenamefont {Walter},
  \citenamefont {Kurpiers}, \citenamefont {Gasparinetti}, \citenamefont
  {Magnard}, \citenamefont {Poto\ifmmode~\check{c}\else \v{c}\fi{}nik},
  \citenamefont {Salath\'e}, \citenamefont {Pechal}, \citenamefont {Mondal},
  \citenamefont {Oppliger}, \citenamefont {Eichler},\ and\ \citenamefont
  {Wallraff}}]{Walter17}%
  \BibitemOpen
  \bibfield  {author} {\bibinfo {author} {\bibfnamefont {T.}~\bibnamefont
  {Walter}}, \bibinfo {author} {\bibfnamefont {P.}~\bibnamefont {Kurpiers}},
  \bibinfo {author} {\bibfnamefont {S.}~\bibnamefont {Gasparinetti}}, \bibinfo
  {author} {\bibfnamefont {P.}~\bibnamefont {Magnard}}, \bibinfo {author}
  {\bibfnamefont {A.}~\bibnamefont {Poto\ifmmode~\check{c}\else
  \v{c}\fi{}nik}}, \bibinfo {author} {\bibfnamefont {Y.}~\bibnamefont
  {Salath\'e}}, \bibinfo {author} {\bibfnamefont {M.}~\bibnamefont {Pechal}},
  \bibinfo {author} {\bibfnamefont {M.}~\bibnamefont {Mondal}}, \bibinfo
  {author} {\bibfnamefont {M.}~\bibnamefont {Oppliger}}, \bibinfo {author}
  {\bibfnamefont {C.}~\bibnamefont {Eichler}}, \ and\ \bibinfo {author}
  {\bibfnamefont {A.}~\bibnamefont {Wallraff}},\ }\href {\doibase
  10.1103/PhysRevApplied.7.054020} {\bibfield  {journal} {\bibinfo  {journal}
  {Phys. Rev. Applied}\ }\textbf {\bibinfo {volume} {7}},\ \bibinfo {pages}
  {054020} (\bibinfo {year} {2017})}\BibitemShut {NoStop}%
\bibitem [{\citenamefont {West}\ \emph {et~al.}(2018)\citenamefont {West},
  \citenamefont {Hensen}, \citenamefont {Jouan}, \citenamefont {Tanttu},
  \citenamefont {Yang}, \citenamefont {Rossi}, \citenamefont {Gonzalez-Zalba},
  \citenamefont {Hudson}, \citenamefont {Morello}, \citenamefont {Reilly} \emph
  {et~al.}}]{west2018gate}%
  \BibitemOpen
  \bibfield  {author} {\bibinfo {author} {\bibfnamefont {A.}~\bibnamefont
  {West}}, \bibinfo {author} {\bibfnamefont {B.}~\bibnamefont {Hensen}},
  \bibinfo {author} {\bibfnamefont {A.}~\bibnamefont {Jouan}}, \bibinfo
  {author} {\bibfnamefont {T.}~\bibnamefont {Tanttu}}, \bibinfo {author}
  {\bibfnamefont {C.}~\bibnamefont {Yang}}, \bibinfo {author} {\bibfnamefont
  {A.}~\bibnamefont {Rossi}}, \bibinfo {author} {\bibfnamefont
  {M.}~\bibnamefont {Gonzalez-Zalba}}, \bibinfo {author} {\bibfnamefont
  {F.}~\bibnamefont {Hudson}}, \bibinfo {author} {\bibfnamefont
  {A.}~\bibnamefont {Morello}}, \bibinfo {author} {\bibfnamefont
  {D.}~\bibnamefont {Reilly}},  \emph {et~al.},\ }\href@noop {} {\bibfield
  {journal} {\bibinfo  {journal} {arXiv:1809.01864}\ } (\bibinfo {year}
  {2018})}\BibitemShut {NoStop}%
\bibitem [{\citenamefont {Aasen}\ \emph {et~al.}(2016)\citenamefont {Aasen},
  \citenamefont {Hell}, \citenamefont {Mishmash}, \citenamefont {Higginbotham},
  \citenamefont {Danon}, \citenamefont {Leijnse}, \citenamefont {Jespersen},
  \citenamefont {Folk}, \citenamefont {Marcus}, \citenamefont {Flensberg},\
  and\ \citenamefont {Alicea}}]{Aasen2016}%
  \BibitemOpen
  \bibfield  {author} {\bibinfo {author} {\bibfnamefont {D.}~\bibnamefont
  {Aasen}}, \bibinfo {author} {\bibfnamefont {M.}~\bibnamefont {Hell}},
  \bibinfo {author} {\bibfnamefont {R.~V.}\ \bibnamefont {Mishmash}}, \bibinfo
  {author} {\bibfnamefont {A.}~\bibnamefont {Higginbotham}}, \bibinfo {author}
  {\bibfnamefont {J.}~\bibnamefont {Danon}}, \bibinfo {author} {\bibfnamefont
  {M.}~\bibnamefont {Leijnse}}, \bibinfo {author} {\bibfnamefont {T.~S.}\
  \bibnamefont {Jespersen}}, \bibinfo {author} {\bibfnamefont {J.~A.}\
  \bibnamefont {Folk}}, \bibinfo {author} {\bibfnamefont {C.~M.}\ \bibnamefont
  {Marcus}}, \bibinfo {author} {\bibfnamefont {K.}~\bibnamefont {Flensberg}}, \
  and\ \bibinfo {author} {\bibfnamefont {J.}~\bibnamefont {Alicea}},\ }\href
  {\doibase 10.1103/PhysRevX.6.031016} {\bibfield  {journal} {\bibinfo
  {journal} {Phys. Rev. X}\ }\textbf {\bibinfo {volume} {6}},\ \bibinfo {pages}
  {031016} (\bibinfo {year} {2016})}\BibitemShut {NoStop}%
\bibitem [{\citenamefont {Didier}\ \emph {et~al.}(2015)\citenamefont {Didier},
  \citenamefont {Bourassa},\ and\ \citenamefont {Blais}}]{Didier2015}%
  \BibitemOpen
  \bibfield  {author} {\bibinfo {author} {\bibfnamefont {N.}~\bibnamefont
  {Didier}}, \bibinfo {author} {\bibfnamefont {J.}~\bibnamefont {Bourassa}}, \
  and\ \bibinfo {author} {\bibfnamefont {A.}~\bibnamefont {Blais}},\
  }\href@noop {} {\bibfield  {journal} {\bibinfo  {journal} {Phys. Rev. Lett.}\
  }\textbf {\bibinfo {volume} {115}},\ \bibinfo {pages} {203601} (\bibinfo
  {year} {2015})}\BibitemShut {NoStop}%
\bibitem [{\citenamefont {Dartiailh}\ \emph {et~al.}(2017)\citenamefont
  {Dartiailh}, \citenamefont {Kontos}, \citenamefont
  {Dou\ifmmode~\mbox{\c{c}}\else \c{c}\fi{}ot},\ and\ \citenamefont
  {Cottet}}]{Dartiailh2017}%
  \BibitemOpen
  \bibfield  {author} {\bibinfo {author} {\bibfnamefont {M.~C.}\ \bibnamefont
  {Dartiailh}}, \bibinfo {author} {\bibfnamefont {T.}~\bibnamefont {Kontos}},
  \bibinfo {author} {\bibfnamefont {B.}~\bibnamefont
  {Dou\ifmmode~\mbox{\c{c}}\else \c{c}\fi{}ot}}, \ and\ \bibinfo {author}
  {\bibfnamefont {A.}~\bibnamefont {Cottet}},\ }\href {\doibase
  10.1103/PhysRevLett.118.126803} {\bibfield  {journal} {\bibinfo  {journal}
  {Phys. Rev. Lett.}\ }\textbf {\bibinfo {volume} {118}},\ \bibinfo {pages}
  {126803} (\bibinfo {year} {2017})}\BibitemShut {NoStop}%
\bibitem [{\citenamefont {Bravyi}(2006)}]{Bravyi2006}%
  \BibitemOpen
  \bibfield  {author} {\bibinfo {author} {\bibfnamefont {S.}~\bibnamefont
  {Bravyi}},\ }\href@noop {} {\bibfield  {journal} {\bibinfo  {journal} {Phys.
  Rev. A}\ }\textbf {\bibinfo {volume} {73}},\ \bibinfo {pages} {042313}
  (\bibinfo {year} {2006})}\BibitemShut {NoStop}%
\bibitem [{\citenamefont {Nickerson}\ \emph {et~al.}(2013)\citenamefont
  {Nickerson}, \citenamefont {Li},\ and\ \citenamefont
  {Benjamin}}]{Nickerson2013}%
  \BibitemOpen
  \bibfield  {author} {\bibinfo {author} {\bibfnamefont {N.~H.}\ \bibnamefont
  {Nickerson}}, \bibinfo {author} {\bibfnamefont {Y.}~\bibnamefont {Li}}, \
  and\ \bibinfo {author} {\bibfnamefont {S.~C.}\ \bibnamefont {Benjamin}},\
  }\href@noop {} {\bibfield  {journal} {\bibinfo  {journal} {Nature Comm.}\
  }\textbf {\bibinfo {volume} {4}},\ \bibinfo {pages} {1756} (\bibinfo {year}
  {2013})}\BibitemShut {NoStop}%
\bibitem [{\citenamefont {Deng}\ \emph {et~al.}(2018)\citenamefont {Deng},
  \citenamefont {Vaitiek\ifmmode~\dot{e}\else \.{e}\fi{}nas}, \citenamefont
  {Prada}, \citenamefont {San-Jose}, \citenamefont {Nyg\aa{}rd}, \citenamefont
  {Krogstrup}, \citenamefont {Aguado},\ and\ \citenamefont
  {Marcus}}]{Deng2018}%
  \BibitemOpen
  \bibfield  {author} {\bibinfo {author} {\bibfnamefont {M.-T.}\ \bibnamefont
  {Deng}}, \bibinfo {author} {\bibfnamefont {S.}~\bibnamefont
  {Vaitiek\ifmmode~\dot{e}\else \.{e}\fi{}nas}}, \bibinfo {author}
  {\bibfnamefont {E.}~\bibnamefont {Prada}}, \bibinfo {author} {\bibfnamefont
  {P.}~\bibnamefont {San-Jose}}, \bibinfo {author} {\bibfnamefont
  {J.}~\bibnamefont {Nyg\aa{}rd}}, \bibinfo {author} {\bibfnamefont
  {P.}~\bibnamefont {Krogstrup}}, \bibinfo {author} {\bibfnamefont
  {R.}~\bibnamefont {Aguado}}, \ and\ \bibinfo {author} {\bibfnamefont {C.~M.}\
  \bibnamefont {Marcus}},\ }\href {\doibase 10.1103/PhysRevB.98.085125}
  {\bibfield  {journal} {\bibinfo  {journal} {Phys. Rev. B}\ }\textbf {\bibinfo
  {volume} {98}},\ \bibinfo {pages} {085125} (\bibinfo {year}
  {2018})}\BibitemShut {NoStop}%
\bibitem [{\citenamefont {Knapp}\ \emph
  {et~al.}(2018{\natexlab{a}})\citenamefont {Knapp}, \citenamefont {Karzig},
  \citenamefont {Lutchyn},\ and\ \citenamefont {Nayak}}]{Knapp2018}%
  \BibitemOpen
  \bibfield  {author} {\bibinfo {author} {\bibfnamefont {C.}~\bibnamefont
  {Knapp}}, \bibinfo {author} {\bibfnamefont {T.}~\bibnamefont {Karzig}},
  \bibinfo {author} {\bibfnamefont {R.~M.}\ \bibnamefont {Lutchyn}}, \ and\
  \bibinfo {author} {\bibfnamefont {C.}~\bibnamefont {Nayak}},\ }\href
  {\doibase 10.1103/PhysRevB.97.125404} {\bibfield  {journal} {\bibinfo
  {journal} {Phys. Rev. B}\ }\textbf {\bibinfo {volume} {97}},\ \bibinfo
  {pages} {125404} (\bibinfo {year} {2018}{\natexlab{a}})}\BibitemShut
  {NoStop}%
\bibitem [{\citenamefont {Fu}(2010)}]{Fu2010}%
  \BibitemOpen
  \bibfield  {author} {\bibinfo {author} {\bibfnamefont {L.}~\bibnamefont
  {Fu}},\ }\href@noop {} {\bibfield  {journal} {\bibinfo  {journal} {Phys. Rev.
  Lett.}\ }\textbf {\bibinfo {volume} {104}},\ \bibinfo {pages} {056402}
  (\bibinfo {year} {2010})}\BibitemShut {NoStop}%
\bibitem [{\citenamefont {Ohm}\ and\ \citenamefont {Hassler}(2015)}]{Ohm2015}%
  \BibitemOpen
  \bibfield  {author} {\bibinfo {author} {\bibfnamefont {C.}~\bibnamefont
  {Ohm}}\ and\ \bibinfo {author} {\bibfnamefont {F.}~\bibnamefont {Hassler}},\
  }\href {\doibase 10.1103/PhysRevB.91.085406} {\bibfield  {journal} {\bibinfo
  {journal} {Phys. Rev. B}\ }\textbf {\bibinfo {volume} {91}},\ \bibinfo
  {pages} {085406} (\bibinfo {year} {2015})}\BibitemShut {NoStop}%
\bibitem [{\citenamefont {Blais}\ \emph {et~al.}(2004)\citenamefont {Blais},
  \citenamefont {Huang}, \citenamefont {Wallraff}, \citenamefont {Girvin},\
  and\ \citenamefont {Schoelkopf}}]{Blais2004}%
  \BibitemOpen
  \bibfield  {author} {\bibinfo {author} {\bibfnamefont {A.}~\bibnamefont
  {Blais}}, \bibinfo {author} {\bibfnamefont {R.-S.}\ \bibnamefont {Huang}},
  \bibinfo {author} {\bibfnamefont {A.}~\bibnamefont {Wallraff}}, \bibinfo
  {author} {\bibfnamefont {S.~M.}\ \bibnamefont {Girvin}}, \ and\ \bibinfo
  {author} {\bibfnamefont {R.~J.}\ \bibnamefont {Schoelkopf}},\ }\href@noop {}
  {\bibfield  {journal} {\bibinfo  {journal} {Phys. Rev. A}\ }\textbf {\bibinfo
  {volume} {69}},\ \bibinfo {pages} {062320} (\bibinfo {year}
  {2004})}\BibitemShut {NoStop}%
\bibitem [{Note1()}]{Note1}%
  \BibitemOpen
  \bibinfo {note} {The analytical expressions for $g_z$ and $\omega _q$
  displayed in~\protect \cref {fig:spectrum} were derived assuming a
  time-independent Hamiltonian. Modulation of the system parameters will lead
  to corrections to these parameters.}\BibitemShut {Stop}%
\bibitem [{\citenamefont {Rainis}\ and\ \citenamefont
  {Loss}(2012)}]{Rainis2012}%
  \BibitemOpen
  \bibfield  {author} {\bibinfo {author} {\bibfnamefont {D.}~\bibnamefont
  {Rainis}}\ and\ \bibinfo {author} {\bibfnamefont {D.}~\bibnamefont {Loss}},\
  }\href {\doibase 10.1103/PhysRevB.85.174533} {\bibfield  {journal} {\bibinfo
  {journal} {Phys. Rev. B}\ }\textbf {\bibinfo {volume} {85}},\ \bibinfo
  {pages} {174533} (\bibinfo {year} {2012})}\BibitemShut {NoStop}%
\bibitem [{\citenamefont {Knapp}\ \emph
  {et~al.}(2018{\natexlab{b}})\citenamefont {Knapp}, \citenamefont {Beverland},
  \citenamefont {Pikulin},\ and\ \citenamefont
  {Karzig}}]{Knapp2018modelingnoiseerror}%
  \BibitemOpen
  \bibfield  {author} {\bibinfo {author} {\bibfnamefont {C.}~\bibnamefont
  {Knapp}}, \bibinfo {author} {\bibfnamefont {M.}~\bibnamefont {Beverland}},
  \bibinfo {author} {\bibfnamefont {D.~I.}\ \bibnamefont {Pikulin}}, \ and\
  \bibinfo {author} {\bibfnamefont {T.}~\bibnamefont {Karzig}},\ }\href
  {\doibase 10.22331/q-2018-09-03-88} {\bibfield  {journal} {\bibinfo
  {journal} {{Quantum}}\ }\textbf {\bibinfo {volume} {2}},\ \bibinfo {pages}
  {88} (\bibinfo {year} {2018}{\natexlab{b}})}\BibitemShut {NoStop}%
\bibitem [{\citenamefont {Royer}\ \emph {et~al.}(2017)\citenamefont {Royer},
  \citenamefont {Grimsmo}, \citenamefont {Didier},\ and\ \citenamefont
  {Blais}}]{Royer2017}%
  \BibitemOpen
  \bibfield  {author} {\bibinfo {author} {\bibfnamefont {B.}~\bibnamefont
  {Royer}}, \bibinfo {author} {\bibfnamefont {A.~L.}\ \bibnamefont {Grimsmo}},
  \bibinfo {author} {\bibfnamefont {N.}~\bibnamefont {Didier}}, \ and\ \bibinfo
  {author} {\bibfnamefont {A.}~\bibnamefont {Blais}},\ }\href {\doibase
  10.22331/q-2017-05-11-11} {\bibfield  {journal} {\bibinfo  {journal}
  {{Quantum}}\ }\textbf {\bibinfo {volume} {1}},\ \bibinfo {pages} {11}
  (\bibinfo {year} {2017})}\BibitemShut {NoStop}%
\bibitem [{\citenamefont {Bravyi}\ \emph {et~al.}(2011)\citenamefont {Bravyi},
  \citenamefont {DiVincenzo},\ and\ \citenamefont
  {Loss}}]{bravyi2011schrieffer}%
  \BibitemOpen
  \bibfield  {author} {\bibinfo {author} {\bibfnamefont {S.}~\bibnamefont
  {Bravyi}}, \bibinfo {author} {\bibfnamefont {D.~P.}\ \bibnamefont
  {DiVincenzo}}, \ and\ \bibinfo {author} {\bibfnamefont {D.}~\bibnamefont
  {Loss}},\ }\href@noop {} {\bibfield  {journal} {\bibinfo  {journal} {Annals
  of physics}\ }\textbf {\bibinfo {volume} {326}},\ \bibinfo {pages} {2793}
  (\bibinfo {year} {2011})}\BibitemShut {NoStop}%
\bibitem [{\citenamefont {C{\'o}rcoles}\ \emph {et~al.}(2015)\citenamefont
  {C{\'o}rcoles}, \citenamefont {Magesan}, \citenamefont {Srinivasan},
  \citenamefont {Cross}, \citenamefont {Steffen}, \citenamefont {Gambetta},\
  and\ \citenamefont {Chow}}]{corcoles2015demonstration}%
  \BibitemOpen
  \bibfield  {author} {\bibinfo {author} {\bibfnamefont {A.~D.}\ \bibnamefont
  {C{\'o}rcoles}}, \bibinfo {author} {\bibfnamefont {E.}~\bibnamefont
  {Magesan}}, \bibinfo {author} {\bibfnamefont {S.~J.}\ \bibnamefont
  {Srinivasan}}, \bibinfo {author} {\bibfnamefont {A.~W.}\ \bibnamefont
  {Cross}}, \bibinfo {author} {\bibfnamefont {M.}~\bibnamefont {Steffen}},
  \bibinfo {author} {\bibfnamefont {J.~M.}\ \bibnamefont {Gambetta}}, \ and\
  \bibinfo {author} {\bibfnamefont {J.~M.}\ \bibnamefont {Chow}},\ }\href@noop
  {} {\bibfield  {journal} {\bibinfo  {journal} {Nature communications}\
  }\textbf {\bibinfo {volume} {6}},\ \bibinfo {pages} {6979} (\bibinfo {year}
  {2015})}\BibitemShut {NoStop}%
\bibitem [{\citenamefont {Hell}\ \emph {et~al.}(2016)\citenamefont {Hell},
  \citenamefont {Danon}, \citenamefont {Flensberg},\ and\ \citenamefont
  {Leijnse}}]{Hell16}%
  \BibitemOpen
  \bibfield  {author} {\bibinfo {author} {\bibfnamefont {M.}~\bibnamefont
  {Hell}}, \bibinfo {author} {\bibfnamefont {J.}~\bibnamefont {Danon}},
  \bibinfo {author} {\bibfnamefont {K.}~\bibnamefont {Flensberg}}, \ and\
  \bibinfo {author} {\bibfnamefont {M.}~\bibnamefont {Leijnse}},\ }\href
  {\doibase 10.1103/PhysRevB.94.035424} {\bibfield  {journal} {\bibinfo
  {journal} {Phys. Rev. B}\ }\textbf {\bibinfo {volume} {94}},\ \bibinfo
  {pages} {035424} (\bibinfo {year} {2016})}\BibitemShut {NoStop}%
\bibitem [{\citenamefont {Cottet}\ \emph {et~al.}(2015)\citenamefont {Cottet},
  \citenamefont {Kontos},\ and\ \citenamefont {Dou\ifmmode~\mbox{\c{c}}\else
  \c{c}\fi{}ot}}]{Cottet15}%
  \BibitemOpen
  \bibfield  {author} {\bibinfo {author} {\bibfnamefont {A.}~\bibnamefont
  {Cottet}}, \bibinfo {author} {\bibfnamefont {T.}~\bibnamefont {Kontos}}, \
  and\ \bibinfo {author} {\bibfnamefont {B.}~\bibnamefont
  {Dou\ifmmode~\mbox{\c{c}}\else \c{c}\fi{}ot}},\ }\href {\doibase
  10.1103/PhysRevB.91.205417} {\bibfield  {journal} {\bibinfo  {journal} {Phys.
  Rev. B}\ }\textbf {\bibinfo {volume} {91}},\ \bibinfo {pages} {205417}
  (\bibinfo {year} {2015})}\BibitemShut {NoStop}%
\bibitem [{\citenamefont {Oreg}\ \emph {et~al.}(2010)\citenamefont {Oreg},
  \citenamefont {Refael},\ and\ \citenamefont {von Oppen}}]{Oreg2010}%
  \BibitemOpen
  \bibfield  {author} {\bibinfo {author} {\bibfnamefont {Y.}~\bibnamefont
  {Oreg}}, \bibinfo {author} {\bibfnamefont {G.}~\bibnamefont {Refael}}, \ and\
  \bibinfo {author} {\bibfnamefont {F.}~\bibnamefont {von Oppen}},\ }\href@noop
  {} {\bibfield  {journal} {\bibinfo  {journal} {Phys. Rev. Lett.}\ }\textbf
  {\bibinfo {volume} {105}},\ \bibinfo {pages} {177002} (\bibinfo {year}
  {2010})}\BibitemShut {NoStop}%
\bibitem [{\citenamefont {Lutchyn}\ \emph {et~al.}(2010)\citenamefont
  {Lutchyn}, \citenamefont {Sau},\ and\ \citenamefont {Sarma}}]{Lutchyn2010}%
  \BibitemOpen
  \bibfield  {author} {\bibinfo {author} {\bibfnamefont {R.~M.}\ \bibnamefont
  {Lutchyn}}, \bibinfo {author} {\bibfnamefont {J.~D.}\ \bibnamefont {Sau}}, \
  and\ \bibinfo {author} {\bibfnamefont {S.~D.}\ \bibnamefont {Sarma}},\
  }\href@noop {} {\bibfield  {journal} {\bibinfo  {journal} {Phys. Rev. Lett.}\
  }\textbf {\bibinfo {volume} {105}},\ \bibinfo {pages} {077001} (\bibinfo
  {year} {2010})}\BibitemShut {NoStop}%
\bibitem [{\citenamefont {Fasth}\ \emph {et~al.}(2005)\citenamefont {Fasth},
  \citenamefont {Fuhrer}, \citenamefont {Bj{\"o}rk},\ and\ \citenamefont
  {Samuelson}}]{fasth2005tunable}%
  \BibitemOpen
  \bibfield  {author} {\bibinfo {author} {\bibfnamefont {C.}~\bibnamefont
  {Fasth}}, \bibinfo {author} {\bibfnamefont {A.}~\bibnamefont {Fuhrer}},
  \bibinfo {author} {\bibfnamefont {M.~T.}\ \bibnamefont {Bj{\"o}rk}}, \ and\
  \bibinfo {author} {\bibfnamefont {L.}~\bibnamefont {Samuelson}},\ }\href@noop
  {} {\bibfield  {journal} {\bibinfo  {journal} {Nano letters}\ }\textbf
  {\bibinfo {volume} {5}},\ \bibinfo {pages} {1487} (\bibinfo {year}
  {2005})}\BibitemShut {NoStop}%
\bibitem [{\citenamefont {Mukhopadhyay}\ \emph {et~al.}(2018)\citenamefont
  {Mukhopadhyay}, \citenamefont {Dehollain}, \citenamefont {Reichl},
  \citenamefont {Wegscheider},\ and\ \citenamefont
  {Vandersypen}}]{Mukhopadhyay20182}%
  \BibitemOpen
  \bibfield  {author} {\bibinfo {author} {\bibfnamefont {U.}~\bibnamefont
  {Mukhopadhyay}}, \bibinfo {author} {\bibfnamefont {J.~P.}\ \bibnamefont
  {Dehollain}}, \bibinfo {author} {\bibfnamefont {C.}~\bibnamefont {Reichl}},
  \bibinfo {author} {\bibfnamefont {W.}~\bibnamefont {Wegscheider}}, \ and\
  \bibinfo {author} {\bibfnamefont {L.~M.}\ \bibnamefont {Vandersypen}},\
  }\href@noop {} {\bibfield  {journal} {\bibinfo  {journal} {Applied Physics
  Letters}\ }\textbf {\bibinfo {volume} {112}},\ \bibinfo {pages} {183505}
  (\bibinfo {year} {2018})}\BibitemShut {NoStop}%
\bibitem [{\citenamefont {Stockklauser}\ \emph {et~al.}(2017)\citenamefont
  {Stockklauser}, \citenamefont {Scarlino}, \citenamefont {Koski},
  \citenamefont {Gasparinetti}, \citenamefont {Andersen}, \citenamefont
  {Reichl}, \citenamefont {Wegscheider}, \citenamefont {Ihn}, \citenamefont
  {Ensslin},\ and\ \citenamefont {Wallraff}}]{Stockklauser2017}%
  \BibitemOpen
  \bibfield  {author} {\bibinfo {author} {\bibfnamefont {A.}~\bibnamefont
  {Stockklauser}}, \bibinfo {author} {\bibfnamefont {P.}~\bibnamefont
  {Scarlino}}, \bibinfo {author} {\bibfnamefont {J.~V.}\ \bibnamefont {Koski}},
  \bibinfo {author} {\bibfnamefont {S.}~\bibnamefont {Gasparinetti}}, \bibinfo
  {author} {\bibfnamefont {C.~K.}\ \bibnamefont {Andersen}}, \bibinfo {author}
  {\bibfnamefont {C.}~\bibnamefont {Reichl}}, \bibinfo {author} {\bibfnamefont
  {W.}~\bibnamefont {Wegscheider}}, \bibinfo {author} {\bibfnamefont
  {T.}~\bibnamefont {Ihn}}, \bibinfo {author} {\bibfnamefont {K.}~\bibnamefont
  {Ensslin}}, \ and\ \bibinfo {author} {\bibfnamefont {A.}~\bibnamefont
  {Wallraff}},\ }\href@noop {} {\bibfield  {journal} {\bibinfo  {journal}
  {Phys. Rev. X}\ }\textbf {\bibinfo {volume} {7}},\ \bibinfo {pages} {011030}
  (\bibinfo {year} {2017})}\BibitemShut {NoStop}%
\bibitem [{\citenamefont {Mi}\ \emph {et~al.}(2018)\citenamefont {Mi},
  \citenamefont {Benito}, \citenamefont {Putz}, \citenamefont {Zajac},
  \citenamefont {Taylor}, \citenamefont {Burkard},\ and\ \citenamefont
  {Petta}}]{Mi2018}%
  \BibitemOpen
  \bibfield  {author} {\bibinfo {author} {\bibfnamefont {X.}~\bibnamefont
  {Mi}}, \bibinfo {author} {\bibfnamefont {M.}~\bibnamefont {Benito}}, \bibinfo
  {author} {\bibfnamefont {S.}~\bibnamefont {Putz}}, \bibinfo {author}
  {\bibfnamefont {D.~M.}\ \bibnamefont {Zajac}}, \bibinfo {author}
  {\bibfnamefont {J.~M.}\ \bibnamefont {Taylor}}, \bibinfo {author}
  {\bibfnamefont {G.}~\bibnamefont {Burkard}}, \ and\ \bibinfo {author}
  {\bibfnamefont {J.~R.}\ \bibnamefont {Petta}},\ }\href@noop {} {\bibfield
  {journal} {\bibinfo  {journal} {Nature}\ }\textbf {\bibinfo {volume} {555}},\
  \bibinfo {pages} {599} (\bibinfo {year} {2018})}\BibitemShut {NoStop}%
\bibitem [{\citenamefont {Samkharadze}\ \emph {et~al.}(2018)\citenamefont
  {Samkharadze}, \citenamefont {Zheng}, \citenamefont {Kalhor}, \citenamefont
  {Brousse}, \citenamefont {Sammak}, \citenamefont {Mendes}, \citenamefont
  {Blais}, \citenamefont {Scappucci},\ and\ \citenamefont
  {Vandersypen}}]{Samkharadze2018}%
  \BibitemOpen
  \bibfield  {author} {\bibinfo {author} {\bibfnamefont {N.}~\bibnamefont
  {Samkharadze}}, \bibinfo {author} {\bibfnamefont {G.}~\bibnamefont {Zheng}},
  \bibinfo {author} {\bibfnamefont {N.}~\bibnamefont {Kalhor}}, \bibinfo
  {author} {\bibfnamefont {D.}~\bibnamefont {Brousse}}, \bibinfo {author}
  {\bibfnamefont {A.}~\bibnamefont {Sammak}}, \bibinfo {author} {\bibfnamefont
  {U.}~\bibnamefont {Mendes}}, \bibinfo {author} {\bibfnamefont
  {A.}~\bibnamefont {Blais}}, \bibinfo {author} {\bibfnamefont
  {G.}~\bibnamefont {Scappucci}}, \ and\ \bibinfo {author} {\bibfnamefont
  {L.}~\bibnamefont {Vandersypen}},\ }\href@noop {} {\bibfield  {journal}
  {\bibinfo  {journal} {Science}\ }\textbf {\bibinfo {volume} {359}},\ \bibinfo
  {pages} {1123} (\bibinfo {year} {2018})}\BibitemShut {NoStop}%
\bibitem [{\citenamefont {Gardiner}\ \emph {et~al.}(2004)\citenamefont
  {Gardiner}, \citenamefont {Zoller},\ and\ \citenamefont
  {Zoller}}]{Gardiner2004}%
  \BibitemOpen
  \bibfield  {author} {\bibinfo {author} {\bibfnamefont {C.}~\bibnamefont
  {Gardiner}}, \bibinfo {author} {\bibfnamefont {P.}~\bibnamefont {Zoller}}, \
  and\ \bibinfo {author} {\bibfnamefont {P.}~\bibnamefont {Zoller}},\
  }\href@noop {} {\emph {\bibinfo {title} {Quantum noise: a handbook of
  Markovian and non-Markovian quantum stochastic methods with applications to
  quantum optics}}},\ Vol.~\bibinfo {volume} {56}\ (\bibinfo  {publisher}
  {Springer Science \& Business Media},\ \bibinfo {year} {2004})\BibitemShut
  {NoStop}%
\end{thebibliography}

%

\end{document}